\documentclass[conference]{IEEEtran}
\IEEEoverridecommandlockouts
\usepackage{cite}
\usepackage{amsmath,amssymb,amsfonts}
\usepackage{algorithmic}
\usepackage{graphicx}
\usepackage{textcomp}
\usepackage{xcolor}
\usepackage[numbers]{natbib}
\usepackage{tabularx}
\usepackage{array}
\usepackage{multirow}
\usepackage{makecell}
\usepackage{float}
\usepackage{placeins}
\usepackage{afterpage}
\usepackage{hyperref}
\usepackage{threeparttable}
\usepackage{balance}
\usepackage{booktabs}
\usepackage{multicol}
\usepackage{url}

\emergencystretch=\maxdimen
\def\BibTeX{{\rm B\kern-.05em{\sc i\kern-.025em b}\kern-.08em
    T\kern-.1667em\lower.7ex\hbox{E}\kern-.125emX}}
\begin{document}

\title{ Are We There Yet? \\A Study of Decentralized Identity Applications \vspace{-0.3cm} }

\author{
    \IEEEauthorblockN{ Daria Schumm, Katharina O. E. Müller, Burkhard Stiller }
    \IEEEauthorblockA{Communication Systems Group (CSG), Department of Informatics (IfI), University of Zurich (UZH), Switzerland \\
    \{schumm, mueller, stiller\}@ifi.uzh.ch }
    \vspace{-0.9cm}
}

\maketitle

\begin{abstract}
    The development of Decentralized Identities (DI) and Self-Sovereign Identities (SSI) has seen significant growth in recent years. 
    This is accompanied by a numerous academic and commercial contributions to the development of principles, standards, and systems. 
    While several comprehensive reviews have been produced, they predominantly focus on academic literature, with few considering grey literature to provide a holistic view of technological advancements. 
    Furthermore, no existing surveys have thoroughly analyzed real-world deployments to understand the barriers to the widespread adoption of decentralized identity models.
    This paper addresses the gap by exploring both academic and grey literature and examining commercial and governmental initiatives, to present a comprehensive landscape of decentralized identity technologies and their adoption in real-world. 
    Additionally, it identifies the practical challenges and limitations that slowdown the transition from centralized to decentralized identity management systems. 
    By shifting the focus from purely technological constraints to real-world deployment issues, this survey identifies the underlying reasons preventing the adoption of decentralized identities despite their evident benefits to the data owner.
\end{abstract}


\section{Introduction}\label{sec:introduction}

In today's digitized world, an increasing number of services, both governmental and commercial, are migrating online \cite{241}. 
This shift towards digital service provision necessitates the development of secure and robust digital identities to ensure safe online interactions. 
However, current centralized and federated identity management approaches are increasingly prone to privacy and security breaches. 
For instance, in April 2024 alone, billions of records of private data were exposed due to cyberattacks on prominent institutions, including a major cancer research center, a popular shopping platform, and one of the leading telecommunications providers in the United States \cite{197}.
These incidents emphasize the vulnerabilities of centralized data storage systems and highlight the urgent need for alternative approaches to personal data management.
A blockchain-based digital identity, particularly Decentralized Identity (DI) and Self-Sovereign Identity (SSI), provides a promising solution to these vulnerabilities. 
DI represents a novel approach to identity management that offers a new perspective on traditional identity management methods. 
Unlike conventional systems that hand over control to centralized databases to store personal data, DI keeps control of data in the hands of its data owners.
Thus, reducing reliance on centralized authorities, thereby enhancing privacy, security, and user autonomy.
Individuals can manage their identities independently, ensuring that their personal information is not susceptible to breaches at a single point of failure.
SSI is a subset of DI, which emphasizes ultimate control over data by the data owner and its disclosure policies.

Despite the significant benefits that such digital identities offer, several challenges delay their widespread adoption in real-world applications. 
Challenges, such as the need for standardization, interoperability and portability, backward compatibility, and governance model, are all significant problems. 
Addressing these steers the development of new technology further.
However, such challenges do not adequately represent the reasons behind the slow adoption of the technology. 
The existing academic literature lacks a complete picture of the DI deployments, focusing on isolated examples and not investigating those used in real-world environments. 
Thus, failing to identify the real-world challenges of DI and SSI.
This paper explores the current state of DI and SSI systems, building on the distinction between the two definitions.
It examines real-world applications and deployments across various domains, providing insights into how these technologies are utilized. 
Additionally, the paper broadens the understanding of the challenges faced by real-world adoption of DI and SSI.  
As a result, gives further directions to understand why a DI and SSI are not widely used in everyday interactions yet.

This survey paper is organized as follows. 
First, Section \ref{sec:background} provides necessary background on the notions of DI and SSI, SSI principles, and technical definitions.
Section \ref{sec:methodology} then outlines the methodology of this survey.
This is followed by a literature review in Section \ref{sec:relatedwork} of the previous survey papers that analyze the state of the art of DI and SSI.
Section \ref{sec:motivation} identifies a research gap in the past literature, and formulates research questions and motivation for this survey.
The most popular standards for DI and SSI are then outlined in Section \ref{sec:standards}, while Section \ref{sec:implementations} provides an overview of system implementations.
Section \ref{sec:realworld} focuses on a selection of implementations available in real-world environments as pilots or complete services, and illustrates the current progress of the DI and SSI adoption.
Following these descriptions of the current landscape, Section \ref{sec:challenges} outlines the adoption challenges of decentralized identities and SSI in real-world scenarios and identifies the gaps beyond technological limitations. 
Finally, Section \ref{sec:conclusion} discusses what challenges should be prioritized to facilitate the transition to the use of an SSI in everyday life.

\section{Background}\label{sec:background}

DI and SSI are key frameworks that reduce reliance on central authorities and enhance user control over personal data, though they differ in principles.


\begin{table*}[t]
    \centering
    \caption{Digital Identity Definitions \& Purposes}
    \vspace{-0.1cm}
    \label{tab:definitions}
    \def\arraystretch{1.5}%
    \footnotesize
    \begin{tabularx}{\textwidth}
    {>{\hsize=0.12\hsize\linewidth=\hsize}X 
     >{\hsize=0.44\hsize\linewidth=\hsize}X 
     >{\hsize=0.44\hsize\linewidth=\hsize}X
    }
        \toprule
        \textbf{Identity} & \textbf{Definition} & \textbf{Purpose} \\
        \midrule
        
        Centralized Identity (CI) & 
        A dynamic set of information owned by third parties and utilized to represent an entity in the digital world. & 
        To facilitate secure digital interactions by enabling identification, authentication, and authorization, allowing an entity to prove their identity electronically and access services. \\
        
        Decentralized Identity (DI) & 
        A technical framework (a service) that addresses the limitations of centralized solutions, emphasizing individual control over data and decentralization of infrastructure. & 
        To enable users to control and manage their data, placing the identity owner at the center of exchange and removing the need for third-party management, while recording decentralized identifiers in the blockchain. \\
        
        Self-Sovereign Identity (SSI) & 
        A digital record is controlled by an end-user, allowing them to manage decisions and have ultimate control over disclosure policies of their personal data, enabling decentralization of infrastructure and governance. & 
        To enhance complete (full) user control and trust in digital interactions, ensuring security, privacy, and minimal disclosure of data. \\
        \bottomrule
    \end{tabularx}
    \vspace{-0.3cm}
\end{table*}

\subsection{DI \& SSI}\label{sec:background_definitions}
    While DI and SSI are frequently used interchangeably, \cite{70} points out that there is a distinction between the two notions. 
    According to the authors, DI is a service that aims to verify user identity and record it in a distributed ledger, such as a blockchain. 
    In contrast, while SSI is a type of DI, the identity is owned by a user without the need to rely on third-party services. 
    \cite{59} argues that designing a digital identity using a blockchain and claiming it is now a self-sovereign identity, is not enough. 
    Every aspect of a system should be outside the control of an organization to avoid pursuing any concept of SSI-as-a-Service \cite{59}. 
    SSI is a subset of the DI and it is possible to have a DI without it being an SSI. 
    \cite{72} classifies SSI as an advanced form of DI, pointing out that in SSI not only are identity attributes controlled by the user but also actions. 

    Despite the ubiquity of definitions for DI and SSI that are available in academic literature, authors commonly only define one but not the other and do not contrast either, leading to confusion over the definition for each. 
    However, while both DI and SSI share the goal of reducing reliance on centralized services and third-party service providers, there is a fine enough line between the definitions to indicate these terms are not interchangeable (Table \ref{tab:definitions} provides a concise summary of the definitions). 
    DI refers to a technical framework that distributes the control and verification of digital identity across stakeholders, minimizing the need for a centralized authority. 
    Its main goal is to remove third-party intermediaries in identity management.
    In contrast, SSI highlights the user’s ultimate control over their digital identity, enabling them to manage and manipulate decisions about the disclosure of information. 
    The concept extends beyond the decentralization aspect, to emphasize individual ownership and control of data, often with a focus on privacy, security, and selective disclosure. 
    While DI provides a broader concept, SSI focuses on the data owner’s autonomy and control over their identity. 
    Additionally, \cite{62} described two further SSI variations: Trust Minimized Identity and Legally-Enabled Self-Sovereign Identity (LESS). 
    The author argues that these two concepts have different goals. 
    The Trust Minimized Identity aims to support the basic human right to identity by providing an identity to those who do not have access to a government-issued identity, while LESS is positioned to have governmental acceptance \cite{62}.

    Blockchain is the central component in DI and SSI.
    It takes the role of a regulatory authority and enables decentralization of power through decentralization of governance and/or infrastructure, making it more difficult to re-purpose data \cite{56, 41, 66}. 
    \cite{49} points out that the role of blockchain in recent SSI implementations is minor and continues to diminish. 
    Nevertheless, the predominant opinion is that blockchain provides a neutral communication channel for participants, guarantees shared control over the identity system and ensures that service providers do not utilize their own private architecture. 
    However, this is not always fulfilled in practice, with many systems controlled by organizations \cite{76}. 
    Furthermore, \cite{66} points out an important distinction between the \textit{decentralization of governance} and the \textit{decentralization of infrastructure}. 
    Since the purpose of SSI is to provide data sovereignty, the decentralization of governance and power over the network should be considered a determining factor. 
    Thus, for an identity system to be considered an SSI, it must have a decentralization of governance, with no single organization or authority having power over system components and participants. 

    To summarize, an SSI is a subset of DI. 
    All SSI are DI but not all DI are SSI. 
    Together in an abstract form, they can be referred to as decentralized identities. 
    DI primarily addresses the decentralization of the identity management system (infrastructure), while SSI emphasizes the ultimate control over data by its data owner (governance). 
    The underlying distinction between DI and SSI is the \textit{decentralization of governance}. 
    For a system to be considered an SSI, every aspect of it should be out of the control of a single organization, software provider, or service.
    Yet, a company can have a decentralized infrastructure (\textit{e.g.}, physical hardware is located in different locations) but centralized governance (\textit{e.g.}, the control over the system).
    For instance, by relying on a native and centralized wallet application for credentials management, a system loses its SSI status and becomes a DI.

\subsection{SSI Principles}\label{sec:background_principles}
    SSI is based on a set of principles defined by Christopher Allen in \cite{61}, summarized in Table \ref{tab:principles_allen}.
    The principles guarantee complete ownership and control over identity, minimal disclosure of the information, system transparency to the user, data neutrality, and portability. 
    The central goal of an SSI is to ensure a single service provider does not hold identities, but identities are neutral, decentralized, and cannot be taken away from the users \cite{59}. 

    A considerable number of authors that discussed DI and SSI systems (e.g., \cite{70, 85, 59, 189, 84}) rely on Cameron’s Laws of Identity \cite{198} for comparison between principles. 
    These principles, outlined in Table \ref{tab:principles_cameroon}, address a digital identity in general rather than a DI or SSI, which might not provide substantial value to the researcher assessing an SSI system. 
    Since there are variations in the SSI principles and definitions, few authors (e.g., \cite{59, 17}) offer an extended set of properties that are built on the original list of Allen \cite{61} and Cameron \cite{198}. 
    The comprehensive list of SSI principles is provided by \cite{17} and summarized in Table \ref{tab:principles_cucko}.
    The authors did not only cover the original set proposed by \cite{61}, but also analyzed additional academic sources to extend the set. 



\begin{table}[h]
    \centering
    \caption{Allen's SSI Principles}
    \vspace{-0.1cm}
    \label{tab:principles_allen}
    \def\arraystretch{1.5}%
    \footnotesize
    \begin{tabularx}{\linewidth}
        {>{\hsize=0.24\hsize\linewidth=\hsize}X 
         >{\hsize=0.76\hsize\linewidth=\hsize}X }
        \toprule
        \textbf{Principle} & \textbf{Definition} \\
        \midrule
        Existence & A user must have an independent existence \\
        Control & A user must control their identity \\
        Access & A user must have access to their own data \\
        Transparency & Systems and algorithms must be transparent \\
        Persistence & An identity must be long-lived \\
        Portability & Information and services must be transportable \\
        Interoperability & An identity should be widely usable \\
        Consent & A user must agree to the use of their identity \\
        Minimization & Disclosure of claims must be minimized \\
        Protection & The rights of a user must be protected \\
        \bottomrule
    \end{tabularx}
    \vspace{-0.5cm}
\end{table}

    
\begin{table*}[h]
    \centering
    \caption{Cameroon's Laws of Identity}
    \vspace{-0.1cm}
    \label{tab:principles_cameroon}
    \def\arraystretch{1.5}%
    \footnotesize
    \begin{tabularx}{\linewidth}
        {>{\hsize=0.27\hsize\linewidth=\hsize}X 
         >{\hsize=0.73\hsize\linewidth=\hsize}X }
        \toprule
        \textbf{Principle} & \textbf{Definition} \\
        \midrule
        User Control and Consent & Users control how their identity is shared and must give consent \\
        Minimal Disclosure & Share only the minimum identity data needed for any interaction \\
        Justifiable Parties & Share identity data only with those who have a valid reason for receiving it \\
        Directed Identity & Use public identifiers for open interactions and private for specific uses \\
        Pluralism & Support multiple technologies and operators for flexibility \\
        Human Integration & Ensure clear, secure communication between users and systems \\
        Consistent Experience & Provide a simple, consistent user experience across different contexts \\
        \bottomrule
    \end{tabularx}
\end{table*}


\begin{table*}[t]
    \centering
    \caption{Cucko's SSI Principles}
    \vspace{-0.1cm}
    \label{tab:principles_cucko}
    \def\arraystretch{1.5}%
    \footnotesize
    \begin{tabularx}{\linewidth}
        {>{\hsize=0.27\hsize\linewidth=\hsize}X 
         >{\hsize=0.73\hsize\linewidth=\hsize}X }
        \toprule
        \textbf{Property} & \textbf{Definition} \\
        \midrule
        Existence and Representation & Entities exist independently and create multiple identities without third-party \\ 
        Decentralization and Autonomy & Entities control their identity data without relying on centralized systems \\
        Ownership and Control & Entities manage and control their own digital identities and data \\
        Privacy and Minimal Disclosure & Entities share only the necessary identity information for interactions \\
        Single Source & Entries are the single source of truth for their identities \\
        Consent & Entities are able to give consent for usage and sharing of their identity data \\
        Security and Protection & Identities and their data are protected through security measures (e.g., encryption) \\
        Verifiability and Authenticity & Identities and data can be verified for authenticity \\
        Accessibility and Availability & Identity data is available and easy to access whenever needed \\
        Recoverability & Identities can be recovered if data is lost or compromised \\
        Usability and User Experience & The system is easy to use and provides a good experience for the entity \\
        Transparency & Entities know how their identity data is being used at all times \\
        Standard & The system follows widely accepted standards for interoperability, portability, persistence \\
        Persistence & Identity data remain valid and accessible for as long as necessary \\
        Portability & Identities and data can be transferred across different systems \\
        Interoperability & Identities work across various systems and platforms \\
        Compatibility & The system works with existing conventional systems \\
        Cost & Cost is minimized for managing and using identities \\
        \bottomrule
    \end{tabularx}
    \vspace{-0.25cm}
\end{table*}

\subsection{SSI Technical Definitions}\label{sec:background_technical}
        \begin{figure*}[p]
            \centering
            \includegraphics[keepaspectratio=true, scale=0.91]{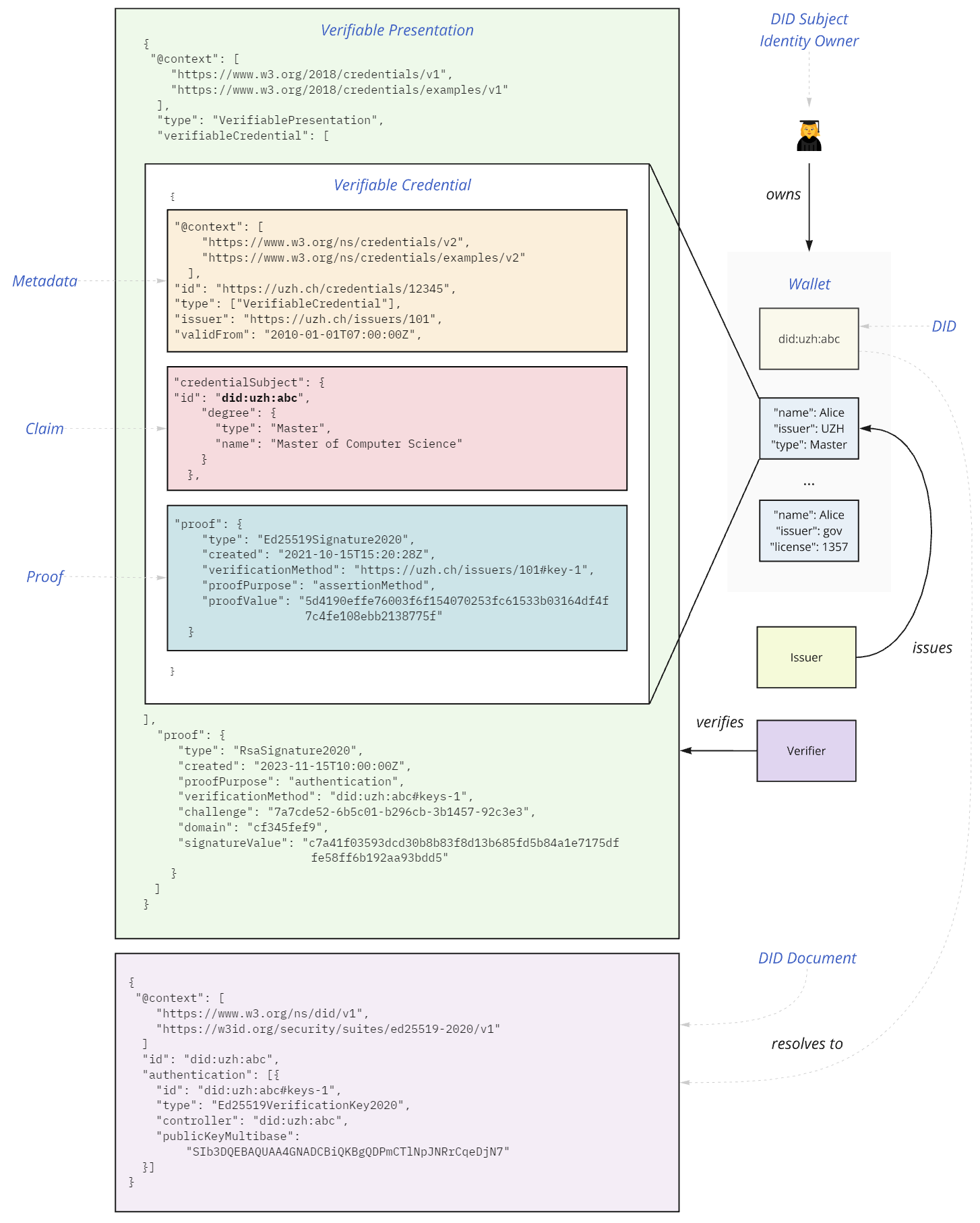}
            \caption{Decentralised Identity with DID, VC and VP}
            \label{fig:ssi_definitions}
        \end{figure*}

    As illustrated in Figure \ref{fig:ssi_definitions}, a DI and SSI contain several components that work together to ensure their functionality. 
    An identity may have several related documents (e.g., academic transcripts), referred to as \textit{credentials}. 
    A credential contains \textit{claims} that are statements about the identity holder \cite{51, 77}. 
    A claim is wrapped in a \textit{verifiable credentials (VC)} or \textit{verifiable presentations (VP)} container. 
    A VC provides cryptographic proof of a claim (e.g. a digital signature), enabling verifiers to check its correctness. 
    A VP constitutes one or more VCs, proves parts of an identity, and authenticates it to the verifier.
    Zero-Knowledge Proof (ZKP) is typically used to disclose a selected part of a claim, known as \textit{selective disclosure}. 
    Selective disclosure is important to ensure compliance with some legal frameworks, such as the General Data Protection Regulation (GDPR) \cite{77}. 
    
    Parties that issue credentials are referred to as \textit{credential providers} or \textit{issuers}. 
    An issuer creates a signature that can verify claims, attest who the issuer was, and guarantee a credential’s correctness \cite{51}. 
    This information is binding and can be authenticated by a \textit{verifier}. 
    Since there may be multiple sources of credentials, the identity system is usually \textit{multi-source}. 
    Multiple credentials can be stored and accessed through a wallet, an application that holds keys for the credential holder and allows identity management. 
    In a multi-source identity system, the identity owner establishes a new relationship with another identity owner (e.g., a service provider) and creates a new \textit{decentralized identifier (DID)} \cite{57}. 
    DID is a unique identifier referencing issuers and identity owners \cite{77}. 
    Since DID creation involves public key exchange, both participants are authenticated \cite{57}. 
    The identity owner is referred to as the \textit{DID subject}. 
    The data that describes the DID subject is referred to as the \textit{DID document} \cite{58}. 
    A \textit{DID controller} can be either a DID subject or another entity that has the right to modify the DID document. 
    To resolve a DID into a meaningful DID document, a \textit{DID method} is necessary. 
    For this, software or hardware, called a \textit{DID resolver}, can be used to retrieve a corresponding DID URL representation \cite{77}.

\section{Methodology}\label{sec:methodology}

For this survey, the selection of literature was divided into two main categories.
The first category looked into past survey papers that focus on the overall landscape of decentralized identities and SSI, provide definitions for both notions, discuss and analyze existing systems, outline principles and requirements for a new type of digital identity, and outline challenges. 
The second category included publications that focus on technical aspects of DI, discuss existing academic and industry-developed systems, and provide some discussion of the real-world adoption of those systems. 
The following exclusion criteria were used: 
\begin{itemize}
    \itemsep0em
    \item Publication language is not English;
    \item Full text of the publication is not accessible;
    \item Focuses on theoretical SSI principles;
    \item Focuses on a particular domain (e.g., healthcare, IoT);
    \item Focuses on a particular operation (e.g., access);
    \item No discussion of any existing DI or SSI systems;
    \item Proposes a new DI or SSI system as a primary contribution;
    \item Provides an identity to devices (e.g., IoT).
\end{itemize}

The initial literature screening was performed across numerous databases, namely IEEE Xplore, Springer Link, Science Direct, ACM Digital Library, and Frontiers in Blockchain. 
The search terms focused on keywords, such as “blockchain” and “identity”. 
Keywords were slightly adjusted to suit the context of the database, such as for Frontiers in Blockchain, the only search term used was “identity” given the database is already focused on the blockchain domain. 
The screening process aimed to capture a comprehensive range of publications from both broader and more niche sources.
Table \ref{tab:search} summarizes the initial databases, the search terms used in each, and the number of results. 


\begin{table*}[!ht]
    \centering
    \caption{Database Search Strings and Number of Results}
    \label{tab:search}
    \def\arraystretch{1.5}
    \footnotesize
    \begin{tabularx}\linewidth
    {>{\hsize=0.25\hsize\linewidth=\hsize}X 
     >{\hsize=0.65\hsize\linewidth=\hsize}X
     >{\hsize=0.10\hsize\linewidth=\hsize}X}
        \toprule
        \textbf{Database}       & \textbf{Search String}                                      & \textbf{Results} \\ 
        \midrule
        IEEE Xplore             & ("All Metadata":blockchain) AND ("Document Title":identity) & 449     \\
        Springer Link           & blockchain AND identity                                     & 257     \\
        Science Direct          & 'blockchain' AND title: 'identity'                          & 95      \\
        ACM Digital Library     & {[}All: blockchain{]} AND {[}Title: "identity"{]}           & 62      \\
        Frontiers in Blockchain & "identity"                                                  & 26      \\
        \bottomrule
    \end{tabularx}
\end{table*}

\begin{figure*}[h]
    \centering
    \includegraphics[width=0.8\textwidth]{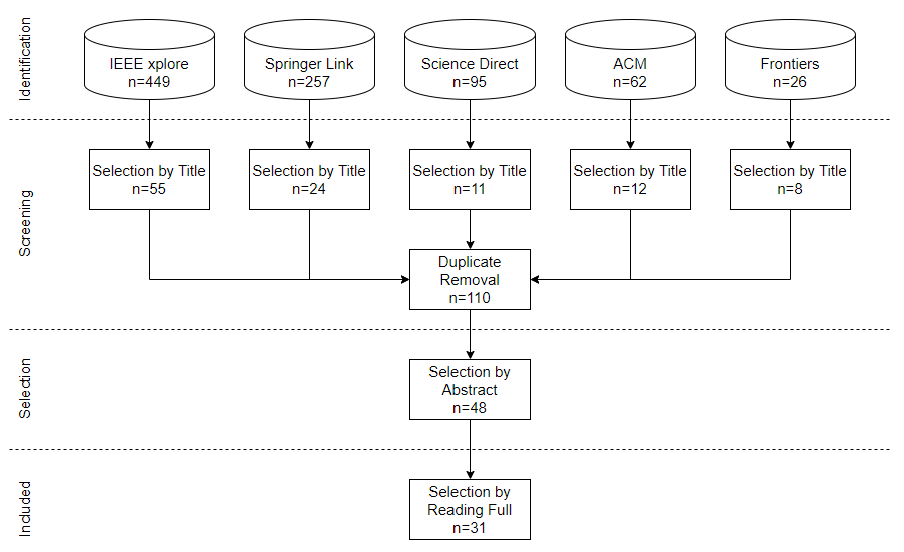}
    \caption{Publications Screening and Selection Process}
    \label{fig:process}
    \vspace{-0.5cm}
\end{figure*}

Following the initial search of each database, the list of results was manually screened and relevant titles were selected. 
The selected titles were then imported into reference management software and duplicates were removed, leaving a total of 110 papers. 
The next step involved reading an abstract of each paper and assessing its suitability to one of the previously mentioned categories (i.e., survey papers and technical papers), with the primary focus on locating relevant survey papers. 
As a result, 48 papers were selected, where 32 papers were allocated to the first category, and 16 to the second category. 
The identified papers were read in full and, as a result, only 31 were selected as a final primary subset.
Figure \ref{fig:process} illustrates the selection process of the primary set of publications. 

Despite providing a substantial starting point, the selected literature from the first iteration was not exhaustive enough to determine the research gap. 
Therefore, more papers were identified using the snowballing technique. 
Subsequent iterations of the research involved looking at (i) the reference lists of the previously identified publications, (ii) the publications that cite the selected papers, and (iii) using the same databases but different keyword searches. 
The keywords used for the subsequent literature identification included “decentralized identity” or “self-sovereign identity” explicitly. 
As a result, an additional 10 publications were discovered for the first category on surveys and 9 for the second category on implementations. 
Section \ref{sec:relatedwork} analyses the most relevant survey papers and identifies the research gap this work addresses. 

Section \ref{sec:realworld} is built predominantly on the analysis of grey literature. 
Since one of the goals is to also provide a survey of real-world implementations or pilots, there was a need to rely more on generic internet searches and screening news articles to identify actual DI deployments. 
First, Google News and specialized news websites (e.g., \url{biometricupdate.com}) were used as a starting point to identify the latest initiatives, with keywords "decentralized identity" and "self-sovereign identity".
Second, after locating projects of interest, any official websites and available documentation were searched to try and identify relevant information, such as the project's purpose, participants, and technical features. 
This information was analyzed and, where possible, conclusions were drawn. 
It is important to note that because the only source of information may be the project itself (e.g., its website or whitepaper), the information drawn may be inadequate to conclude and may require additional practical testing of the system. 
\section{Related Work}\label{sec:relatedwork}

The further iterative selection of publications resulted in the 16 most relevant academic surveys based on their contribution to analyzing the state of the art. 
As a result, this work examines those surveys, identifies a research gap and future research directions. 
Each survey was analyzed based on the following parameters, which are also used in the subsequent discussion, as shown in Tables \ref{tab:implementations_ssi} and \ref{tab:implementations_non_ssi}. 
First, whether authors addressed the distinction between DI and SSI, and categorized existing systems as such. 
Second, whether the authors identified and analyzed DI and SSI systems based on technical and infrastructural considerations. 
Technical features include considerations such as (i) interoperability and portability of decentralized identifiers and identity data, (ii) data minimization, (iii) whether a system is a DI or SSI, (iv) type of blockchain, and (v) the use of DIDs and VCs. 
Infrastructure properties consider whether a system (i) utilizes governmental ID for user verification and/or (ii) biometric data for authentication, (iii) reliance on any centralized elements, (iv) whether it is open source, (v) whether it offers a native mobile application, and (vi) if it is currently active. 
Third, whether authors identified and discussed real-world DI or SSI systems that are currently in use or previously had a pilot program. 
The real-world system is defined as a system that is not commonly discussed in academic literature, but is popular in the industry, deployed in a specific domain (e.g., government services), and involves a substantial number of users (e.g., one million). 
Such a system is used in a real-world environment and supports interactions between users and other stakeholders on an everyday basis. 
Fourth, whether the authors address the DI and SSI adoption challenges. 
Table \ref{tab:surveys} summarizes the literature review of the surveys.

\begin{table*}[t]
    \caption{Related Work Summary}
    \vspace{-0.1cm}
    \label{tab:surveys}
    \def\arraystretch{1.5}%
    \footnotesize
    \begin{tabularx}{\textwidth}
        {>{\hsize=0.05\hsize\linewidth=\hsize}X 
         >{\hsize=0.10\hsize\linewidth=\hsize}X 
         >{\hsize=0.65\hsize\linewidth=\hsize}X 
         >{\hsize=0.15\hsize\linewidth=\hsize}X 
         >{\hsize=0.25\hsize\linewidth=\hsize}X 
         >{\hsize=0.25\hsize\linewidth=\hsize}X 
         >{\hsize=0.20\hsize\linewidth=\hsize}X 
         >{\hsize=0.20\hsize\linewidth=\hsize}X }
        \toprule
        \textbf{Ref.} & \textbf{Year} & \textbf{Authors} & \textbf{DI or SSI} & \textbf{Technical Analysis} & \textbf{Infrastructure Analysis} & \textbf{Real-World Systems} & \textbf{Adoption Challenges} \\
        \midrule
        \cite{70}   & 2018 & Dunphy and Petitcolas              & Yes & Limited & Limited & No & Yes \\
        \cite{85}   & 2019 & Haddouti and Kettani               & No & Very Limited & Very Limited & No & No \\
        \cite{94}   & 2019 & Ferdous, Chowdhury and Alassafi    & No & Yes & Yes & No & No \\
        \cite{59}   & 2020 & Satybaldy, Nowostawki and Ellingsen& No & Yes & Yes & No & Yes \\
        \cite{72}   & 2020 & Dib and Toumi                      & No & Very Limited  & Limited & No & Limited \\
        \cite{87}   & 2020 & Kuperberg                          & No & Very Limited & Yes & No & Limited \\
        \cite{189}  & 2020 & Liu et al.                         & No & No & No & No & No \\
        \cite{190}  & 2020 & Kaneriya and Patel                 & No & Limited & Very Limited & No & No \\
        \cite{44}   & 2021 & Ghaffari et al.                    & No & Very Limited & No & No & No \\
        \cite{84}   & 2021 & Soltani, Nguyen and An             & No & Very Limited & Very Limited & No & Very Limited \\
        \cite{186}  & 2021 & Zaeem et al.                       & Yes & Yes & No & No & No \\
        \cite{6}    & 2022 & Bai et al.                         & No & Limited & No & No & Limited \\
        \cite{79}   & 2022 & Ahmed et al.                       & Yes & Yes & Limited & No & No \\
        \cite{191}  & 2022 & Rathee and Singh                   & No & No & Very Limited & No & No \\
        \cite{192}  & 2023 & Tan, Chi and Lam                   & No & No & No & No & No \\
        \cite{193}  & 2024 & Buttar et al.                      & No & No & No & No & Yes \\
        \bottomrule
    \end{tabularx}
\end{table*}

\cite{70} conducted one of the first surveys on blockchain-based identity management, defining decentralized trusted identity (DTI) and SSI, and evaluating three systems (uPort, Sovrin, and ShoCard) based on Cameron’s laws of identity \cite{198}. 
Despite discussing some technical and infrastructural aspects, the survey lacks coverage of critical issues like portability and interoperability and does not categorize the systems as DTI or SSI. 
While considering real-world use cases and practical challenges, the analysis is limited in the number of systems and examples of DI or SSI real-world use cases.

Similarly to \cite{70}, \cite{85} analyzed uPort, Sovrin, and ShoCard using Cameron’s Laws of Identity, aiming to provide a more detailed discussion. 
The authors used the terms DI and SSI interchangeably and described the components of each system. 
However, the analysis is limited, lacking clarity on how the comparison was performed, evidence for conclusions, and a comprehensive overview of the challenges. 
Additionally, the authors did not consider real-world use cases and challenges of real-life adoption. 

\cite{94} conducted a comprehensive survey on SSI, providing a formal concept and mathematical model for SSI based on SSI principles (e.g., existence, availability, interoperability). 
The authors analyzed uPort, Sovrin, Jolocom, and Blockcerts based on SSI properties, using technical documentation and whitepapers. 
However, the paper lacks a discussion on the distinction of SSI from DI and does not address challenges and real-world adoption examples, while emphasizing the need for real-life use cases to understand the usefulness and applicability of SSI. 

\cite{59} propose an evaluation framework for SSI systems, based on Cameron’s Laws of Identity, with an additional requirement for usability. 
Authors evaluate Sovrin, uPort, ShoCard, Civic, and Blockstack through their documentation and whitepapers, providing a comprehensive technical and infrastructure analysis, and SSI principles. 
While the paper discusses challenges such as centralization, user interactions, and economic barriers, it does not examine real-world adoption examples.  

\cite{72} expanded the analysis of decentralized identities and included IDchainZ, EverID, LifeID, and SelfKey alongside uPort, Sovrin, and ShoCard, but without differentiating between DI and SSI. 
Authors evaluated these systems using SSI principles but provided misleading labeling of infrastructure features as non-functional assessments. 
The paper provides limited infrastructure comparison and lacks in-depth technical discussion, making it unclear how conclusions on each SSI principle were reached. 
Challenges of SSI systems, including adoption, are briefly discussed without depth, and real-world system considerations are not mentioned.   

\cite{87} provides a comprehensive survey on blockchain-based identity management (IdM) systems, including both market-available and academic proposals, but without differentiating between DI and SSI. 
The survey emphasizes real-world considerations and bridges the gap between market research and academic perspective, although it mixes non-system initiatives into systems analysis (e.g., Decentralized Identity Foundation (DIF)). 
The author introduces criteria for enterprise requirements divided into compliance and liability, user experience, technology, and operations and integration. 
Despite not providing an evaluation of systems used in real-world settings, the survey concludes that Blockpass IDN, Civic, ShoCard, Sovrin, and uPort are the most suitable for enterprise IdM based on maturity and adoption levels. 

\cite{189} compare uPort, Sovrin, and ShoCard based on Cameron’s Laws of Identity, similar to \cite{70, 85}, but lack technical and infrastructural analysis. 
The authors review literature focused on authentication, privacy, and trust, but without connecting this review to the analysis of the three systems. 
The paper does not explore real-world systems discussion or adoption challenges, and the discussion of challenges lacks depth. 

\cite{190} study various SSI implementations, including Sovrin, uPort, EverID, LifeID, Sora, and SelfKey, with limited technical and infrastructural analysis. 
The discussion of challenges is brief and lacks consideration of real-world adoption.
The paper offers a high-level overview of integration use cases in SSI (e.g., enrollment and usage of identity), but these are generic, lacking technical details and specific challenges.  

\cite{44} provide an overview and taxonomy of blockchain-based identity and access management (IAM) systems, including limited technical discussion of SSI systems such as uPort, Sovrin, ShoCard, SelfKey, and Identity Overlay Network (ION). 
The authors cover blockchain-based access control solutions and outline challenges in blockchain-based IdM systems. 
However, the paper uses the term SSI to refer to both DI and SSI systems, does not include a discussion on real-world deployment, and mentions user experience as a future research direction that requires further exploration to provide a meaningful challenge.  

\cite{84} provide another survey of SSI, focusing on existing platforms, regulatory frameworks, and building blocks.
Authors discuss the challenges in traditional identity management, and the motivation for SSI, and provide definitions and summaries based on Cameron’s Laws of Identity, but do not distinguish between DI and SSI. 
The paper briefly reviews several SSI systems, such as Blockstack, Civic, and Sovrin, but offers limited technical and infrastructural analysis, lacks descriptions and does not cover real-world deployments or in-depth adoption challenges. 

\cite{186} conducted a survey analyzing commercial and academic work on SSI, uniquely providing functional requirements (FRs) for SSI systems and comparing existing offerings based on these FRs. 
The authors categorize some systems based on the distinctions between DI and SSI as defined by \cite{70}, but lack reasoning for this categorization and do not apply it universally. 
The survey does not discuss real-world systems or address challenges.

\cite{84} summarizes digital identity evolution, including motivation for an SSI, and they introduce blockchain-based architecture. 
The authors do not distinguish between DI and SSI notions and use them interchangeably. 
The survey briefly discusses technical features of the existing systems, such as ShoCard, uPort, and Sovrin, but does not address infrastructure considerations in detail. 
The paper addresses real-world challenges such as user experience, regulatory compliance, and privacy conflicts but lacks a detailed outlook on real-world SSI deployments and a strong foundation for discussing adoption challenges.

\cite{79} provide a comprehensive survey of academic and market blockchain-based IdM systems, distinguishing between DI and SSI definitions. 
The authors provide a technical overview of some systems (uPort, Sovrin, ShoCard, Jolocom, Civic, etc.), but lack detailed infrastructure discussion.
The paper reviews a substantial number of blockchain-based IdMs from academic literature not covered in any previous works, compares commercial systems based on SSI principles and infrastructure, and outlines technological and blockchain-specific challenges, but lacks discussion of broader implementation and real-world deployment issues.  

\cite{191} investigate blockchain-based IdM systems, highlighting how blockchain addresses the challenges of traditional IdM approaches. 
The authors identified several DI and SSI systems (e.g., uPort, Sovrin, ShoCard), providing infrastructure information for each, but lacking in-depth technical considerations. 
The main contribution of the paper is the inclusion of research project initiatives using blockchain-based IdMs, however, these projects are inherently research-focused and lack insights into real-world adoption. 
Moreover, the study does not discuss technological or real-world adoption challenges. 

\cite{192} provide an overview of the SSI concept, as well as the concept of self-sovereignty within digital identity, but do not distinguish between DI and the SSI definitions and use the two interchangeably. 
The authors identify DI systems, such as uPort, Sovrin, and ShoCard, but lack technical and infrastructural considerations for each. 
Additionally, real-world deployments are not mentioned and the paper does not address adoption challenges.  

\cite{193} focus on blockchain-based IdM systems, using concepts of DI and SSI interchangeably. 
The authors investigate the adoption challenges, including user experience, infrastructure, integration, governance, and standards issues, providing a comprehensive overview. 
Additionally, the authors discuss potential use cases and benefits of DI for specific scenarios. 
However, the survey lacks analysis of real-world systems, as well as not provide technical and infrastructural discussion of existing systems.

\section{Motivation \& Contribution}\label{sec:motivation}

Over the past few years, the development of DI and SSI, and research in this space, has been growing steadily. 
Few authors have produced outstanding reviews of the field, largely relying on academic conferences and journal papers.
Few studies have considered grey literature to provide a comprehensive overview of new technology development.
However, no survey has analyzed the real-world deployments to understand the challenges that slow adoption. 
Although such an approach may be sufficient from a purely academic perspective, it does not identify problems that delay the shift toward a DI model in real-world conditions. 
Nevertheless, this is a prerequisite to facilitate such a change and bring better data control into everyday life patterns. 
Despite the technological advances in the area and a growing number of commercial initiatives, there are still reasons why such a beneficial approach to a user fails to transition centralized identity management mechanisms toward decentralization. 
Instead of addressing the question of adoption, the existing works on DI and SSI are excessively concerned with the technological limitations of the systems and underlying principles of SSI. 
To progress toward decentralized identity management in the real world, there is a need to first ask what has been done so far and why it has not yet worked. 
To answer these questions, there is a need to establish a clearer view of the technology landscape that is not built exclusively on academic work but includes a wide range of commercial and governmental initiatives. 
This survey answers three research questions: 
\begin{itemize}
    \itemsep0em
    \item RQ1: What is the current state of the art in DI and SSI systems?
    \item RQ2: What is currently used in a real-world environment? 
    \item RQ3: What are the challenges to a wide adoption of an SSI technology? 
\end{itemize}

Therefore, the contributions of this survey are threefold: (i) discussing standards and providing the comprehensive analysis, categorization, and technical and infrastructural highlights of the current DI and SSI systems (Section \ref{sec:implementations}, answering RQ1), (ii) looking beyond academic literature and supplementing the state of the art by outlining real-world DI and SSI system deployments (Section \ref{sec:realworld}, answering RQ2), and (iii) outlining challenges for real-world SSI system adoption and complementing those with detailed technological, societal, and economic challenges (Section \ref{sec:challenges}, answering RQ3).

\section{Standards}\label{sec:standards}

Despite the already existing variety of implementations, standards in DI and SSI are only starting to be developed and actively integrated. 
The leading players in defining standards for DI are the World Wide Web Consortium (W3C) and the Decentralized Identity Foundation (DIF), with smaller organizations offering further standardization. 
Both working groups do not emphasize the SSI approach but address DI as a general concept.
Table \ref{tab:standards} summarizes popular standards. 


\begin{table*}[!ht]
    \centering
    \caption{Decentralised Identity and SSI Standards}
    \vspace{-0.1cm}
    \label{tab:standards}
    \def\arraystretch{1.5}%
    \footnotesize
    \begin{threeparttable}
        \begin{tabularx}{\textwidth}
        {>{\hsize=0.15\hsize\linewidth=\hsize}X 
         >{\hsize=0.30\hsize\linewidth=\hsize}X
         >{\hsize=0.05\hsize\linewidth=\hsize}X
         >{\hsize=0.45\hsize\linewidth=\hsize}X
         >{\hsize=0.07\hsize\linewidth=\hsize}X}
            \toprule
            \textbf{Name} & \textbf{Author} & \textbf{Year} & \textbf{Scope} & \textbf{Status} \\
            \midrule
            DID & W3C & 2022 & Definition of new decentralized and verifiable identifiers & Active \\
            VC & W3C & 2024 & Express secure, privacy-preserving and verifiable credentials & Draft \\
            DID Resolution & W3C, Credentials Community Group & 2024 & Resolving DIDs and de-referencing DID URLs & Draft \\
            Universal Resolver & DIF & 2022 & Resolve DIDs into DID Documents & Active \\
            DIDComm & DIF & 2023 & Communication protocol for DID & Draft \\
            ERC-725 & ERC-725 Alliance & 2020\tnote{a} & Ethereum smart contracts accounts & Active \\
            \bottomrule
        \end{tabularx}
        \vspace*{5px}
        \begin{tablenotes}
            \item[a] First release 
        \end{tablenotes}
    \end{threeparttable}
    \vspace{-0.3cm}
\end{table*}

\subsection{W3C}
    W3C is a non-profit organization that develops standards and guidelines for the web and promotes accessible and interoperable technologies.
    The following standards were developed by the Decentralized Identifier Working Group \cite{239}, Verifiable Credentials Working Group \cite{238}, and Credentials Community Group \cite{237}.

    \subsubsection{Decentralized Identifiers (DID)}
        The main contribution of the W3C toward DI standardization is the architecture, data model, and representation of DIDs. 
        The recommendation, outlined in \cite{120}, provides definitions, syntax, and data structure for DID, DID subject, and DID document, as well as outlines the DI architecture and operations. 
        The purpose of the recommendation is to supply DI providers with a common approach to specifying DIDs and enabling interoperability and portability across the systems. 
        Interoperability and portability are the fundamental principles of SSI. 
        The former refers to the seamless exchange of credentials between entities despite the differences in the underlying technologies, while the latter means the transfer of identity data between different systems and devices. 
        Standardization aims at a syntactical level of interoperability that guarantees the implementations rely on the same data format (e.g., JSON). 
        Differences in implementations and lack of standards adoption are one of the primary reasons that the syntactical interoperability layer is not easily achieved. 
        The W3C recommendation provides a unified approach to DIDs, improving interoperability and portability and ensuring identity data can be used between different participants. 
        This standard is supported by the US Department of Homeland Security \cite{160}. 
    
    \subsubsection{Verifiable Credentials (VC)}
        The standard for VCs is less stable than standards for DIDs, with the current recommendation continuing to be a draft, and a refined version available in 2024. 
        VC aims to address the same problem of interoperability by providing a unified data structure for defining credentials. 
        
    \subsubsection{DID Resolution}
        DID Resolution does not have a finalized standard, but a draft version by the W3C Credentials Community Group.
        The draft specifies the process of obtaining a DID Document using a DID, that contains information such as public keys and service endpoints that are necessary to facilitate an interaction with the subject. 
        The DID Resolution is performed by the DID Resolver, which is a hardware or software that takes a DID as an input and returns a DID Document as an output. 

\subsection{DIF}
    DIF is a foundation developing open and accessible standards for DI.
    It is composed of multiple working groups that address identifiers, DID authentication, DID communication, claims and credentials, data storage, wallet security, and cryptographic protocols. 
    The organization includes multiple members and partners, that represent universities, commercial companies, and governmental organizations \cite{121}. 

    \subsubsection{Universal Resolver}\label{sec:standards_resolver}
        Interoperability was addressed from a different perspective by the DIF working groups, with the main contribution being the Universal Resolver. 
        Due to interoperability limitations, it may not be possible to make one identity from one blockchain interact with another blockchain. 
        Thus, there is a need for a Universal Resolver in the middle, which can resolve a specific DID method.  
        Since there are many DID methods available, there is a need for a process to revolve one DID method to someone who wants to communicate with that DID \cite{77}. 
        The Universal Resolver can be run locally or requested through HTTP and can convert a DID into a DID document, get cryptographic keys, service endpoints, and metadata \cite{22}. 
        However, the current architecture of a universal resolver is centralized and works as a trusted service between the client and blockchain \cite{22}. 
    
    \subsubsection{DIDComm}
        DID Communication (DIDComm) is a decentralized technique for addressing and relaying messages by establishing an authenticated communication channel \cite{77}. 
        It provides a bi-directional communication channel between two participants who know each other’s DIDs \cite{125}. 
        The solution defines how communication mechanisms work in application-level protocols and workflows while preserving trust \cite{124}. 
        According to \cite{77}, it is one of the most adopted interoperable communications between two parties today. 

\subsection{ERC-725}
    ERC-725 is not mentioned in the academic literature. 
    Nevertheless, it has strong community support and proposes a standard for blockchain-based SSI \cite{120}. 
    It is a standard for creating, publishing, and managing DI through a smart contract on an EVM-based blockchain. 
    The standard defines a proxy smart contract that is controlled by multiple parties and other smart contracts, and outlines the standard for adding and removing claims \cite{127}. 
    The ERC-725 targets an SSI rather than a general DI concept, allowing identity owners to manage their own identity data.

\section{Implementations}\label{sec:implementations}
In recent years, the world of DI and SSI has been rapidly evolving, with numerous organizations and researchers contributing to the development of platforms and standards. 
This section offers an overview of the key players in the DI and SSI landscape and examines systems that provide identities to individuals.

Since many sources use the terms DI and SSI interchangeably, there is an inconsistency in opinions about whether a specific system is an SSI or not. 
Based on reviewed literature (discussed in Section \ref{sec:methodology}), some popular systems (e.g., ShoCard, Jolocom) were classified by the authors as DI rather than SSI. 
However, as noted previously in Section \ref{sec:background_definitions}, a DI does not necessarily incorporate important SSI principles, such as interoperability and portability. 
DI may still rely on centralized elements for identity verification \cite{79}. 
Meanwhile for a system to be considered an SSI, every aspect of the organization that provides an SSI system should be out of the control of the organization, thus, achieving decentralization of governance \cite{59}. 
Therefore, this section classifies current academic and industry identity providers as SSI and non-SSI, based on how other authors identified those systems. 
A system was marked as non-SSI if at least one academic source classified it as such or the system does not identify as SSI but as a DI. 
The purpose of this is to illustrate the continuous ambiguity in the current understanding of what makes a system an SSI.

\subsection{SSI}\label{sec:implementations_ssi}
    This section discusses a few of the most cited academic papers on SSI systems, such as uPort, Sovrin, SelfKey, Civic, and LifeID, and extends the analysis with less known systems that have the potential to grow and be utilize in real-world environments, such Truvity, Gataca, Privado ID. 
    Table \ref{tab:implementations_ssi} summarizes the infrastructure and technical features of the SSI systems. 
    
    
\begin{table*}[!ht]
    \centering
    \caption{Infrastructure and Technical Features of SSI Systems}
    \vspace{-0.1cm}
    \label{tab:implementations_ssi}
    \def\arraystretch{1.5}
    \footnotesize
    \begin{threeparttable}
        \begin{tabularx}{\textwidth}
        {>{\hsize=0.40\hsize\linewidth=\hsize}X 
         >{\hsize=0.18\hsize\linewidth=\hsize}X
         >{\hsize=0.25\hsize\linewidth=\hsize}X 
         >{\hsize=0.25\hsize\linewidth=\hsize}X 
         >{\hsize=0.25\hsize\linewidth=\hsize}X 
         >{\hsize=0.25\hsize\linewidth=\hsize}X 
         >{\hsize=0.25\hsize\linewidth=\hsize}X 
         >{\hsize=0.30\hsize\linewidth=\hsize}X 
         >{\hsize=0.20\hsize\linewidth=\hsize}X 
         >{\hsize=0.30\hsize\linewidth=\hsize}X 
         >{\hsize=0.35\hsize\linewidth=\hsize}X 
         >{\hsize=0.15\hsize\linewidth=\hsize}X }
            \toprule
            \textbf{Name} & \textbf{Gov. ID\tnote{a}} & \textbf{Biom.\tnote{b}} & \textbf{Central. El.\tnote{c}} & \textbf{Open Source} & \textbf{Mobile App} & \textbf{Active} & \textbf{Interop.\tnote{d} \& Port.\tnote{e}} & \textbf{Data Min.\tnote{f}} & \textbf{Pub.\tnote{g} or Priv.\tnote{h}} & \textbf{Blockchain} & \textbf{DID VC} \\
            \midrule
            uPort & No & No & Yes & Yes & Native & No & No & No & Public & Ethereum & Yes \\
            Sovrin & No & No & No & Yes & Native & Yes & No & Yes & Private & Hyperledger & Yes \\
            SelfKey & No & No & No & Yes & Native & Yes & ? & Yes & Public & Ethereum & Yes \\
            Civic & Yes & Yes & Yes & ? & Native & Yes & No & Yes & Public & Ethereum & ? \\
            LifeID & No & Yes & ? & Yes & ? & No & ? & Yes & Public & Ethereum & ? \\
            Truvity & No & No & ? & No & Native & Yes & ? & ? & ? & ? & Yes \\
            Gataca & Yes & No & ? & No & Native & Yes & ? & Yes & Any & Agnostic & Yes \\
            Privado ID & Yes & Yes & ? & Yes & Native & Yes & ? & ? & Public & Polygon & Yes \\
            \bottomrule
        \end{tabularx}
        \vspace*{-0.2cm}
        \begin{tablenotes}
            \begin{multicols}{2}
                \item[a] Government ID
                \item[b] Biometrics
                \item[c] Centralised Elements
                \item[d] Interoperable 
                \item[e] Portable
                \item[f] Data Minimisation
                \item[g] Public
                \item[h] Private
            \end{multicols}
        \end{tablenotes}
    \end{threeparttable}
    \vspace{-0.5cm}
\end{table*}

    \subsubsection{uPort}
        uPort is a highly cited SSI framework in the academic literature. 
        Currently, it no longer exists as a standalone system, but was split into subsequent DI projects - Serto and Veramo. 
        Many recent papers (e.g., \cite{187}) fail to acknowledge the non-existence of the system and continue to list it as one of the widely available frameworks. 
        Despite this, it continues to dominate the SSI discussion and, thus, it is worth mentioning its main features. 
        
        uPort was developed as an open-source SSI provider, based on the public Ethereum blockchain and three smart contracts, namely controller contract, proxy contract, and registry contract \cite{77, 79, 94, 105}.
        The purpose of the smart contracts was to (i) provide identity owners with access control and recovery mechanisms, (ii) enable linking uPort identifiers with private keys, and (iii) link between uPort identifiers and the off-chain storage, which helps to locate the required DID \cite{77, 94, 105}.
        Additionally, the framework utilized the InterPlanetary File System (IPFS) to store identity information and related DID documents, making it more scalable \cite{59, 77, 94}. 
        The system also utilized four centralized elements for (i) communication between mobile applications and any decentralized application compatible with uPort, (ii) removing the need for a new user to pay Ethereum gas fees when creating an account, (iii) providing an interface to communicate with the Ethereum network and (iv) with IPFS storage \cite{59, 105}.
        Identity owners were provided with a mobile application to manage keys and personal data \cite{59}. 
        
        \cite{44, 72, 82, 105} point out that uPort identities lack portability and interoperability since they are rooted on-chain in the Ethereum network. 
        Additionally, \cite{72} mentions that uPort identities do not provide unlinkability and malicious nodes can trace all the activities of a single identity, compromising user privacy. 
        uPort did not address data minimization but allowed a user to selectively discard attributes, enabling users to permanently remove some information such as criminal records \cite{59, 94}. 
        \cite{94} states that the operational costs may present a barrier to wide-scale adoption of such a system due to its reliance on the public Ethereum network and transactions.  
        Additionally, scalability remained an issue for uPort regardless of the use of IPFS \cite{105}.
        Despite being the most discussed and active SSI system, \cite{87} points out that the operational website and production-level services were almost nonexistent. 
        Moreover, the mobile application, libraries, and services are all depreciated now \cite{79}. 
        Serto and Veramo projects inherited uPort functionalities, with Veramo being considered the new uPort. 

    \subsubsection{Serto}
        Serto aims to provide enterprises with a DI system that includes a mobile wallet and credential management capabilities. 
        There is no indication whether Serto is an SSI. 
        Based on \cite{106}, Serto was promised to become commercially available in 2020.
        However, at the time of writing, Serto website and related articles about its ecosystem were not available. 

    \subsubsection{Veramo}
        Veramo is an open-source framework that provides a modular API for SSI, that enables users to create and manage DI, as well as verify credentials \cite{107}. 
        The framework aims to improve uPort limitations by providing standardized (based on W3C and DIF recommendations discussed in Section \ref{sec:standards}) and interoperable DI systems. 
        The system operated through a Veramo DID Agent, which provides a gateway to the framework \cite{108}. 
        According to \cite{109}, Veramo uses Ethereum to store DIDs on-chain, while \cite{102} mentions that Veramo is blockchain agnostic and supports few DID methods.
        Additionally, \cite{102} points out that Veramo is still in the beta stage and functionalities are not properly implemented.
        Moreover, scarcely any academic discussion is available on Veramo. 

    \subsubsection{Sovrin}\label{sec:implementations_sovrin}
        Sovrin is a non-profit US-based foundation that provides DI and is based on the public permissioned Hyperledger Indy blockchain \cite{84, 110}. 
        \cite{104} identifies Sovrin as an Identity as a Service (dIaaS) platform. 
        Anyone can use the network, but only pre-approved parties, usually trusted institutions and organizations, can participate in the consensus and have the write access \cite{13, 79}.
        Write access is necessary to create VCs within the Sovrin network \cite{148}. 
        The reliance on pre-selected nodes forces users to rely on a middleware between users and blockchain \cite{59, 72}.
        Today Sovrin Foundation includes large companies, such as IBM and Cisco, as their data stewards \cite{144}. 
        Currently, Sovrin is operating three networks with 4 to 25 writing nodes each \cite{111}, that run a novel consensus mechanism called Plenum \cite{94, 113}.
        Plenum, in turn, is based on the Redundant Byzantine Fault Tolerance (RBFT) protocol and was developed by the Hyperledger Indy foundation \cite{173}. 
        Four ledgers are combined to run the Plenum consensus, one of which holds all identity records \cite{113}. 
        
        Users can interact with the ledger through the mobile application or website, which works as a Sovrin client and allows them to create, update, manage, and share the identity data \cite{94}. 
        An identity owner enters into a formal agreement with the Sovrin Foundation before the identity is created \cite{87}. 
        Data is held by the owner in their digital wallet on the edge or a third-party cloud, which are regarded as agents for secure communication between system participants (e.g. users and issuers) \cite{105, 114, 94}. 
        However, \cite{59} points out that all personal data is stored on the user device and \cite{72} states that no claim is registered on the blockchain. 
        Additionally, Sovrin enables key recovery based on initially nominated trustees \cite{72}. 
        Similarly to Hyperledger Indy, Sovrin provides users with attribute-based credentials and data minimization through selective disclosure \cite{44, 82, 79}. 
        However, \cite{44, 85, 105} mention that portability and interoperability are not yet supported. 
        Meanwhile, \cite{94} argues that Sovrin has a portability feature, because it relies on a specific standard to represent an identity, but will lose this feature once the corresponding ledger ceases to exist.  
        \cite{84, 105} marked Sovrin as an open-source platform, but did not highlight that it is open-source because it operates on Hyperledger Indy, while Sovrin itself provides a layer on top of it \cite{112}.
        The documentation for the Sovrin layer is limited; it was last updated in 2018, and the public repository page is not exceptionally active. 
    
    \subsubsection{SelfKey}
        SelfKey is an open-source SSI framework that is based on the private instance of the Ethereum blockchain \cite{83}. 
        It is based on its own wallet application and utilizes native tokens \cite{44, 87}. 
        According to \cite{159}, SelfKey overlooks user control and consent and identity data persistence. 
        However, SelfKey was modified in 2023.
        The updated version incorporated more W3C standards, adoption of privacy-preserving Know Your Customer (KYC) functionality for credential issuance, and transformation toward a Decentralized Autonomous Organization (DAO) governance model based on crypto-economic incentives \cite{118}. 

    \subsubsection{Civic}\label{sec:implementation_civic}
        According to \cite{87}, Civic is one of the largest blockchain-based identity services based on a market segment.
        Civic supports Ethereum and Solana blockchains by providing a library adapted for each, but also claims to be available on other blockchains (e.g., Avalanche, Polygon) \cite{175}.  
        \cite{79, 84} point out that a closed-source mobile application, which does not support data import or synchronization, is utilized for identity data storage and management.
        However, Civic documentation claims the system is wallet agnostic \cite{176}. 
        Users can share their identity data selectively \cite{79}. 
        Blockchain stores the hashes of the identity data as ERC20 tokens \cite{59}, which is Ethereum standard for fungible tokens \cite{174}. 
        
        \cite{59, 72} point out that Civic is not completely decentralized and user identities may depend on the Civic existence, and thus may not be persistent over time. 
        Additionally, it requires legal documents to verify the identity \cite{72}. 
        \cite{79} points out that this system lacks portability as it relies on authentication authorities. 
        Similarly, \cite{159} argues that the system does not support portability, as well as lacks persistence because it relies on a third party. 
        \cite{104} argues that Civic is a blockchain-based KYC system, while \cite{102} classifies it as an SSI. 
        However, Civic documentation does not claim the system to be an SSI. 

    \subsubsection{LifeID}
        LifeID was open source and based on Ethereum but is currently an inactive project \cite{72, 83, 84}. 
        It incorporated principles of SSI, enabled data minimization through the use of Zero-Knowledge Proof (ZKP), and stored data on the user’s device \cite{79}. 
        Additionally, the framework used biometric authentication instead of passwords \cite{79}. 
        \cite{159} points out the system had significant issues with privacy and security but did not discuss those issues further. 

    \subsubsection{Truvity}
        The private and commercial software company offers an SSI API for developers to integrate SSI identity management into their businesses. 
        The information on the website is limited, with developer documentation covering high-level basics of SSI functionality and a whitepaper inaccessible to the general public without providing an email address \cite{195}.
        The source code is closed source and thus, there is no way to verify company claims about their system. 
        The system relies on W3C DID and VC, as well as utilizes DIDComm. 
        The website points out that the API provides an "easy-to-deploy" cloud platform and lacks information on blockchain infrastructure, which raises questions about the decentralization of the system \cite{194, 195}. 
        Additionally, the scarce documentation and information on the website do not refer to the system as a DI. 

    \subsubsection{Gataca}\label{sec:implementation_gataca}
        Gataca is a private and commercial company based in Spain and provides users with a DI system. 
        The official website uses DI and SSI definitions interchangeably but lacks any technical documentation or a whitepaper. 
        Users can use a native wallet to manage their identity data or an online platform that enables integration of DI with the user's application or website \cite{202}.
        However, \cite{216} point out they were unable to set up a mobile wallet.
        
    \subsubsection{Privado ID (Polygon ID)}
        The system aims to address increasing identity theft and fraud, preserve the privacy of personal information, and mitigate the risks of misinformation generated by artificial intelligence (AI) \cite{203}.
        Privado ID is open-source and utilizes W3C standards and biometric verification. 
        The initiative enables various companies and organizations to contribute to their ecosystem by developing applications for credentials issuance and verification \cite{204}. 
        Dock (Section \ref{sec:implementation_dock}) and Civic (Section \ref{sec:implementation_civic}) are part of the ecosystem. 
        The system allows users to utilize a Privado ID wallet or a compatible wallet offered by companies in their ecosystem \cite{205}. 
        Moreover, ecosystem applications offer governmental ID verification services and the use of biometric data.
        Since the initiative enables users to connect other applications supported by the Privado ID ecosystem, it is difficult to conclude whether the system uses centralized elements. 
        Interoperability and portability of personal data and credentials similarly depend on the ecosystem elements utilized by a user. 
        
\subsection{Non-SSI}\label{sec:implementations_non_ssi}
    This section discusses systems that are non-SSI, or DTI, such as PingOne Neo, Jolocom, Stacks, IDchainZ, Dock, Midy, Worldcoin, and Spherity. 
    Table \ref{tab:implementations_non_ssi} summarises the infrastructure and technical features of the non-SSI systems. 
    
    
\begin{table*}[!ht]
    \centering
    \caption{Infrastructure and Technical Features of Non-SSI Systems}
    \vspace{-0.1cm}
    \label{tab:implementations_non_ssi}
    \def\arraystretch{1.5}
    \footnotesize
    \begin{threeparttable}
        \begin{tabularx}{\textwidth}
        {>{\hsize=0.47\hsize\linewidth=\hsize}X 
         >{\hsize=0.18\hsize\linewidth=\hsize}X
         >{\hsize=0.25\hsize\linewidth=\hsize}X 
         >{\hsize=0.25\hsize\linewidth=\hsize}X 
         >{\hsize=0.25\hsize\linewidth=\hsize}X 
         >{\hsize=0.25\hsize\linewidth=\hsize}X 
         >{\hsize=0.25\hsize\linewidth=\hsize}X 
         >{\hsize=0.30\hsize\linewidth=\hsize}X 
         >{\hsize=0.20\hsize\linewidth=\hsize}X 
         >{\hsize=0.30\hsize\linewidth=\hsize}X 
         >{\hsize=0.35\hsize\linewidth=\hsize}X 
         >{\hsize=0.15\hsize\linewidth=\hsize}X }
            \toprule
            \textbf{Name} & \textbf{Gov. ID\tnote{a}} & \textbf{Biom.\tnote{b}} & \textbf{Central. El.\tnote{c}} & \textbf{Open Source} & \textbf{Mobile App} & \textbf{Active} & \textbf{Interop.\tnote{d} \& Port.\tnote{e}} & \textbf{Data Min.\tnote{f}} & \textbf{Pub.\tnote{g} or Priv.\tnote{h}} & \textbf{Blockchain} & \textbf{DID VC} \\
            \midrule
            PingOne Neo & Yes & Yes & Yes & No & Any & Yes & No & No & Public & Bitcoin & Yes \\
            Jolocom & No & No & ? & Yes & Native & No & ? & No & Public & Ethereum & Yes \\
            Stacks & No & No & ? & Yes & - & Yes\tnote{i} & No & No & Public & Bitcoin & No \\
            IDchainZ & Yes & No & Yes & No & Native & No & No & No & ? & ? & No \\
            Blockcerts & No & No & ? & Yes & Native & Yes & ? & No & Public & Bitcoin & Yes \\
            Dock & No & No & Yes & Yes & Native & Yes & ? & Yes & Public & ? & Yes \\
            Midy & Yes & Yes & ? & No & Native & Yes & No & Yes & ? & ? & Yes \\
            Worldcoin & Yes & Yes & Yes & Partially & Native & Yes & No & Yes & Public & Ethereum & ? \\
            Spherity & Yes & ? & ? & No & Native & Yes & ? & ? & ? & ? & ? \\
            \bottomrule
        \end{tabularx}
        \vspace*{-0.2cm}
        \begin{tablenotes}
            \begin{multicols}{2}
                \item[a] Government ID
                \item[b] Biometrics
                \item[c] Centralised Elements
                \item[d] Interoperable 
                \item[e] Portable
                \item[f] Data Minimisation
                \item[g] Public
                \item[h] Private
                \item[i] As Layer-2 Blockchain Solution
            \end{multicols}
        \end{tablenotes}
    \end{threeparttable}
    \vspace{-0.5cm}
\end{table*}

    \subsubsection{PingOne Neo (ShoCard)}
        Even though ShoCard was acquired by PingIdentity in 2020 and no longer exists as a standalone framework, many academic papers still discuss it as the original ShoCard.  
        Some sources refer to the ShoCard as an SSI but \cite{83} regard it as Decentralized Trusted Identity (DTI) instead and \cite{6, 79} refer to it as blockchain-based DI. 
        PingOne Neo, provided by PingIdentity, labels their system as DI, stating that sometimes it is referred to as SSI \cite{115}. 
        Thus, the system does not make a distinction between the two notions. 
        
        ShoCard and its successor are not open source \cite{83}. 
        Both are based on blockchain and utilize biometric data, such as facial recognition, for user identification \cite{54}. 
        Government identification is required before identity is issued \cite{59, 115}. 
        ShoCard was designed to support VCs, blockchain-based authentication and data management \cite{44}.
        According to \cite{44, 79}, data minimization was not supported. 
        Likewise, PingOne Neo does not mention data minimization or ZKP in the documentation on credential presentation \cite{117}.  
        Any mobile application can be used to download the issued credentials \cite{85, 116}. 
        However, ShoCard did not support the export of the data to any secondary or on-device storage \cite{87}. 
        Moreover, the ShoCard framework did not provide necessary data privacy, portability, and persistence of the identity data \cite{159}. 
        Additionally, \cite{59, 72} point out that it was partially centralized as it relied on intermediary servers between users and relying parties, which created an uncertainty about the persistence of the existence of the identities \cite{59}. 
    
    \subsubsection{Jolocom}
        Jolocom is another widely cited digital identity provider, but within an agreement on whether it is an SSI framework.
        \cite{79} states that similarly to ShoCard, Jolocom is a DTI, not an SSI. 
        Contradictory opinion expressed by \cite{102} that classifies Jolocom as an SSI.         

        The open-source framework was based on Ethereum and IPFS, as well as utilized DID and VC standards specified by W3C \cite{84, 79}. 
        A registry smart contract stored DID hashes \cite{79, 94}, while IPFS stored DID Documents. 
        The framework did not consider data minimization \cite{94}. 
        \cite{44, 94} point out that Jolocom had similar functionalities to uPort, but utilizes different data structures.  
        Users can use the native mobile application to interact, create, manage, and share their identity data \cite{79, 94}. 
        At the moment of writing, the Jolocom website and whitepaper are no longer accessible and the public repositories have not been active for years.
        The Jolocom SmartWallet appears to be the most updated repository. 
    
    \subsubsection{Stacks (Blockstack)}
        Blockstack is an open-source naming and storage platform that aims to redesign the naming system \cite{44, 84}. 
        \cite{70} mentions that Blockstack extends the Namecoin framework and provides a linkage between a public key and a human-readable identifier, thus achieving a decentralized public key infrastructure (PKI). 
        Several papers (e.g., \cite{83, 84, 44, 59, 79, 104}) list Blockstack as one of the DI approaches, while also describing it as a decentralized naming and storage platform. 
        However, Blockstack never positioned itself as an identity management system, with its whitepaper outlining the goals of the system as decentralized naming and discovery service, and decentralized storage, and does not mention DI \cite{208}. 
        The system did offer Blockstack ID, but at the time of writing the corresponding repository was deprecated. 
        Additionally, the name Blockstack is no longer used and is currently the Stack, which is now the Bitcoin Layer 2 solution \cite{119}.

    \subsubsection{IDchainZ}
        IDchainZ is a proof of concept prototype that is not completely implemented \cite{82, 83, 72}. 
        The project is not currently active. 
        \cite{83, 79} categorized this system as DTI, not an SSI. 
        The concept relies on the use of blockchain and provides a mechanism to exchange identity and KYC documents between participants (e.g., user and verifier). 
        There is no indication of which blockchain (e.g., private, public) the system relies on, as well as no discussion of data minimization, interoperability, and portability of data. 
        The system utilizes government IDs to verify users and relies on at least one centralized element, Attribute Exchange Platform (the IDchainZ platform itself), that stores identity information and matches the requests against it \cite{209}.
        \cite{72} points out that identities are highly dependent on the IDchainZ platform and, thus, not persistent. 
        The system offers users a wallet application to manage identity data \cite{209}. 

    \subsubsection{Blockcerts}
        Blockcerts is an open-source credential system that utilizes the Ethereum or Bitcoin blockchain network to store and verify the cryptographic hash of a digital certificate \cite{94}.
        A certificate can represent information such as a civil record, academic credentials, and licenses \cite{207}.  
        \cite{94} points out the system is not a "fully-fledged" SSI, does not use a data minimization approach, and incurs high operating costs. 
        Additionally, the authors did not provide a conclusion on whether Blockcerts certificates are interoperable and portable \cite{94}. 
        However, according to \cite{207}, the framework is aligned with W3C DID and VC standards. 
        The system relies on the open-source native wallet for holding, viewing, and verifying credentials. 
        Blockcerts require an issuer to be added to the system before it can issue credentials. 
        At the time of writing, the Blockcerts repository is active but does not have much activity. 

    
    \subsubsection{Dock}\label{sec:implementation_dock}
        Dock is an open-source DI system that utilizes a native blockchain designed for DI, and supports VCs and DIDs \cite{150, 151, 152}. 
        The company is funded by Web3 Foundation \cite{156}. 
        Dock uses a native wallet to receive, manage, and store credentials \cite{153}. 
        It provides an abstract library for interaction with blockchain and enables issuers and verifiers to make use of its functionalities without the need for technical knowledge. 
        The company targets organizations and individuals to provide them with a user-friendly way to issue, manage, verify credentials, and address the problem of document forgery \cite{154}.
        The official website does not claim to provide an SSI system but continuously refers to Dock as a DI. 
        Moreover, Dock states that SSI and DI terms are interchangeable and that SSI is based on three pillars that are blockchain, VCs, and DIDs \cite{130}. 
        Based on the discussion in the previous sections, it is evident that the notion of SSI involves many more variables than purely technical considerations. 
        There are currently no academic papers discussing this system in detail. 

    \subsubsection{Midy (Evernym)}
        Evernym is a for-profit software company that specializes in SSI, and is currently owned by Avast \cite{232}, the leading corporation in digital security and privacy. 
        Originally, the company founded the Sovrin Foundation (discussed in Section \ref{sec:implementations_sovrin}) and donated code to Hyperledger Indy and Aries \cite{158}. 
        It is not clear whether Evernym offered a separate identity framework from Sovrin, but \cite{159} outlined Evernym and Sovrin frameworks separately. 
        Additionally, \cite{159} points out that Evernym did not comply with user control and interoperability principles.
    
        Currently, Midy is the new face of Everynym. 
        It is a closed-source DI system that aims to provide quicker proof of a unique human instead of CAPTCHA and is based on W3C DID and VC standards. 
        The system requires a user to scan a government-issued identity document and record a video of themself, before allowing them to create a digital credential and cryptographic pseudonym \cite{131}. 
        The pseudonym created from the credential is different for each service the user engages with \cite{131}. 
        The whitepaper does not specify whether and which blockchain Midy uses. 
        
    \subsubsection{Midy}
        Midy is the new face of Evernym, according to the Evernym official website \cite{178}.
        It is a closed-source DI platform that aims to provide quicker proof of a unique human instead of CAPTCHA, and is based on DID and VC standards \cite{131}.
        However, it is not clear whether it provides a DI or SSI. 
        The platform requires a user to scan a government-issued identity document and record a video of himself, before enabling them to convert them into a digital credential. 
        The cryptographic pseudonym is generated from the credential data. 
        The document is scanned once but can be used in multiple proofs. 
        The pseudonym created from the credential is different for each service the user engages with \cite{131}. 
        However, the whitepaper does not address the linkability issue within the same service, meaning that the behavior of the user could still be tracked within a single service provider. 
    
    \subsubsection{Worldcoin}\label{sec:implementations_worldcoin}
        Worldcoin aims to provide a globally inclusive identity, and financial framework, and promote a global economy for all \cite{132}. 
        The system relies on the Ethereum blockchain but utilizes a centralized database to verify whether a person was already registered by using a hash of their iris scan, and to perform verification of ZKPs (Gent, 2023) \cite{129}. 
        Altruistic in theory, the actual implementation received significant criticism ranging from the use of centralized hardware to a lack of linkage between individual biometrics and their WorldID and unethical collection of biometric data from the first million users \cite{129, 134}.
        \cite{134} point out that the project is surrounded by misinformation, with the main concern being the privacy of the biometric data and motivation to collect “most data for this AI-driven economy”. 
        Currently, despite being a US-based company, Worldcoin is not available in the US, China, Turkey, and Sudan due to local regulatory constraints. 
        Spain has also banned the use of Worldcoin iris scan \cite{135}. 
        It is noteworthy that the Worldcoin whitepaper does not contain the keywords “decentralized identity” nor “self-sovereign identity”, but references to “privacy-preserving identity network” and identity claims are reported to be verified in a decentralized manner \cite{132}. 
        Therefore, it is difficult to conclude whether Worldcoin provides a DI or a traditional approach to digital identity without further investigation of its architecture. 

    \subsubsection{Spherity}
        According to \cite{157}, Spherity is a “pioneer in decentralized identity management software”. 
        The company provides multiple applications, such as (i) wallet software for managing digital identities, (ii) a product passport, and (iii) supply chain management software.
        The company does not provide a whitepaper nor technical documentation to analyze what features offered systems have, which leads to the impossibility of concluding certain aspects of the system, such as data minimization, presence of centralized elements, and use of W3C standards. 
        The wallet supports the management of DI and can be integrated with different ecosystems, such as Gaia-X \cite{210, 233}. 
        Additionally, Spherity is one of the Sovrin steward nodes \cite{213}.

\section{Real-World Environment}\label{sec:realworld}
There are countless attempts to bring a decentralized approach to identity management in real-world scenarios. 
Yet, real-world adoption remains the most challenging. 
Not only do deployments frequently face technological limitations (discussed in Section \ref{sec:challenges_tech}), but also business constraints, societal conflicts, and scarce discussion and reporting of the adoption progress and result.  
At the time of writing, government-backed initiatives dominate the deployment of DI and SSI. 
This section outlines a few examples in governmental services, education, and travel, including Zug ID, QuarkID, German ID, Mebuku Ground, Kiva, Buthan National Digital Identity, Spanish Universities Pilot, IATA Travel Pass, and WEF Known Traveler Digital ID. 
The Table \ref{tab:real_world} summarizes discussed initiatives. 


\begin{table*}[!ht]
    \centering
    \caption{Real-World Decentralised Identity and SSI Implementations}
    \vspace{-0.1cm}
    \label{tab:real_world}
    \def\arraystretch{1.5}
    \footnotesize
    \begin{threeparttable}
        \begin{tabularx}{\textwidth}
        {>{\hsize=0.70\hsize\linewidth=\hsize}X 
         >{\hsize=0.25\hsize\linewidth=\hsize}X 
         >{\hsize=0.25\hsize\linewidth=\hsize}X 
         >{\hsize=0.13\hsize\linewidth=\hsize}X 
         >{\hsize=0.13\hsize\linewidth=\hsize}X 
         >{\hsize=0.15\hsize\linewidth=\hsize}X 
         >{\hsize=0.40\hsize\linewidth=\hsize}X 
         >{\hsize=0.13\hsize\linewidth=\hsize}X 
         >{\hsize=0.13\hsize\linewidth=\hsize}X 
         >{\hsize=0.15\hsize\linewidth=\hsize}X }
            \toprule
            \textbf{Name} & \textbf{Industry} & \textbf{Location} & \textbf{Year} & \textbf{Active} & \textbf{DI\tnote{a} or SSI} & \textbf{Infrastructure} & \textbf{DID + VC} & \textbf{Gov. ID\tnote{b}} & \textbf{Open Source} \\
            \midrule
            Zug ID & Government & Switzerland & 2017 & No & SSI & uPort & Yes & Yes & Yes \\
            QuarkID & Government & Argentina & 2024 & Yes & SSI & ? & Yes & Yes & No \\
            IDunion & Various & Germany & 2022 & Yes & SSI & Hyperledger Aries Hyperledger Indy & Yes & ? & ? \\
            Mebuku Ground & Government & Japan & 2022 & Yes & SSI & ? & ? & Yes & No \\
            Kiva Protocol & Government & Sierra Leone & 2019 & No & DI & Hyperledger Aries Hyperledger Indy Hyperledger Ursa & Yes & No & Yes \\
            Buthan National Digital Identity & Government & Buthan & 2021 & Yes & SSI & Hyperledger Indy & Yes & Yes & ? \\
            Spanish Universities Pilot & Education & Spain & 2024 & Yes & SSI & EBSI\tnote{c} & Yes & ? & No \\
            EQAR & Education & EU & 2021 & Yes & SSI & EBSI & Yes & ? & Yes \\
            IATA Travel Pass & Travel & International & 2021 & No & SSI & Sovrin & ? & No & No \\
            WEF Known Traveler Digital ID & Travel & Canada Netherlands & 2023 & Yes & DI & Hyperledger Indy & ? & Yes & No \\
            \bottomrule
        \end{tabularx}
        \vspace*{5px}
        \begin{tablenotes}
            \item[a] Decentralized Identity 
            \item[b] Government ID 
            \item[c] European Blockchain Service Infrastructure
        \end{tablenotes}
    \end{threeparttable}
    \vspace{-0.3cm}
\end{table*}

\subsection{Government Services}
    There are many attempts to adopt DI and SSI in governmental services. 
    However, there is little discussion and reporting, leading to confusion about whether a particular initiative is a simple digital identity, a blockchain-based DI or SSI. 
    Frequently, governments that are behind such initiatives do not assign their system to one of the categories, but either use definitions interchangeably or do not use them at all, resorting to some generic terms (e.g., DI). 
    
    \subsubsection{ZugID}
        ZugID was initiated in 2017 in the canton of Zug, Switzerland, and provided residents with a DI to access governmental services and participate in e-voting \cite{162}.
        The application was based on uPort and utilized a native wallet for identity management. 
        The pilot was completed in 2020, but there were no reports on whether uPort's deprecation the following year impacted ZugID and no conclusion on the program's results. 
        Currently, the new Zug eID platform provided by a Swiss company Procivis \cite{164} provides DI and SSI, based on DIDs and VCs, as well as native wallet applications for identity management.
        However, not only is Procivis naturally business-oriented, but the product lacks clear benefit to the canton's residents. 
        Furthermore, reliance on ready-to-use and user-friendly systems provided by a private company introduces a significant dependency for the government on that software provider \cite{7}.

    \subsubsection{QuarkID}
        QuarkID is an open-source SSI that was recently developed in collaboration with government agencies, and adopted by the city of Buenos Aires, Argentina, in 2024. 
        At the time of writing, there are more than 2500 users \cite{199}. 
        The main aim is to provide citizens with access and control over legal documents \cite{145}.
        The framework utilizes W3C DID and VC standards, as well as DIDComm for secure communication between system participants \cite{181}. 
        The system is blockchain agnostic and supports Ethereum, Polygon, and Rootstock networks. 
        There is very little information available on the system's performance in everyday interactions. 
        However, QuarkID provides improved educational resources to the users and emphasizes data sovereignty throughout the system. 

    \subsubsection{Germany eID \& IDunion}
        Although Germany has a clear initiative to develop an SSI-based national ID and achieve decentralization, the information is scattered and lacks a coherent summary and description of the technology used. 
        Multiple sources, such as \cite{212}, point out that the German identity card program relies on a decentralized architecture, but lacks further discussion. 
        Other sources, such as \cite{220} and \cite{221}, categorize German eID as a user-centering IDM, meaning that the system gives some control over identity data to the user, but does not employ a complete decentralized infrastructure and utilizes centralized elements (e.g., eID server). 
        It is difficult to gather information on the technical aspect of the current system since numerous official sources were not accessible at the time of writing and no new publications analyzed it. 
        Thus, one is limited in providing a conclusion on the current German ID system in this paper. 
        However, based on this research, the decentralized approach to identity management is in development and largely driven by the Electronic Identification and Trust Services (eIDAS) 2.0 regulation, affecting not only Germany but other European Union (EU) countries. 
        The current eIDAS version aims to provide guidelines on interoperable digital identities that are recognizable across borders \cite{219}.
        The updated regulation is planned to come into effect in 2026 and is expected to give legal recognition to blockchains, thus, creating an opportunity for DI \cite{222}. 
        
        At the time of writing, the decentralization of Germany's national ID involves numerous projects, with IDUnion \cite{218} and Lissi wallet \cite{217} being the most important contributions. 
        IDUnion relies on Hyperledger Aries \cite{234} and Indy \cite{235}, and aims to create an SSI ecosystem, ensuring the interoperability of identity data within the EU, selective disclosure, and utilizing existing W3C and DIF technical standards \cite{218}. 
        The project has run multiple pilots in various sectors, such as government, education, and finances \cite{223}.
        Two identity wallets are supported: Lissi and esatus. 
        Lissi was established as a startup company and is supported by the German government. 
        The project provides two solutions: a wallet application to manage credentials and a connector to the European Digital Identity Wallets to enable seamless connection for various use cases \cite{217}. 
        
    \subsubsection{Mebuku Ground}
        Mebuku Ground is a local, government-backed system, initiated in 2022 by the business community of Maebashi, Japan, and established by Mebuku Ground Inc. 
        The system includes 57 local businesses, educational institutes, and the Maebashi local government. 
        Initially created as an identity service provider, the company emphasizes data governance, the right of users to their own data, and marks itself as SSI. 
        The system utilizes governmentally issued ID to issue VCs and anonymous credentials and enables selective disclosure of credential attributes\cite{103}.
        However, according to \cite{103}, for legal and security reasons, private information can be revealed to trace the user. 
        The authors do not provide further details about this process. 
        Additionally, based on the architecture, Mebuku Ground provides a trusted centralized element that maintains the record of user permissions and revocations \cite{103}.
        Therefore, raising a concern over the persistence of a user identity. 
        Furthermore, there is a lack of information about whether the system utilizes blockchain and follows W3C standards.
        In 2024 the system was also adopted in Omura city \cite{211}. 
    
    \subsubsection{Kiva Protocol}
        The Kiva Protocol was the first African DI platform, created in partnership with the Sierra Leone government and a non-profit organization that facilitates crowdfunding loans to unbanked and underbanked populations.
        The primary goal of the platform was to provide an inclusive identity to those who lacked formal identification. 
        Trust anchors, such as government bodies and microfinance institutions, issue VCs, ensuring that citizens can build a credit history even in the absence of a national credit bureau \cite{188}.
        The provided DI utilized w3C DID and VC standards, involved an open-source developer community, and did not require a governmental ID to enable KYC between the users and verifiers to satisfy regulatory compliance \cite{224}. 
        The project was discontinued in 2022 without clear reason \cite{225}. 
 
    \subsubsection{Buthan National Digital Identity (NDI)}
        Buthan NDI is another example of SSI real-world adoption by a government body alongside various organizations, such as financial institutions, education providers, and transportation authorities. 
        Identities are based on government-issued ID and biometric data \cite{227}.
        The system enables participants to issue, exchange, and verify VCs through a native NDI Wallet and utilizing DID and VC and enables data minimization \cite{226}.
        The system employs two registries. 
        First, the Trust Registry records public DIDs of trusted organizations.
        Second, the Verifiable Data Registry, based on Hyperledger Indy, stores issuer public keys, DID documents, DID schema, and metadata \cite{226}. 
        The details about architecture are not available and it is not clear whether the Trust Registry is a centralized element. 

\subsection{Education}
    There are few use cases of utilizing decentralized approaches to identity management in the education sector, with most being small-scale projects still in the development phase.
    The use of DI or SSI to issue and verify education certificates is a popular example of real-world deployment. 
    Since education-related projects are smaller in scale than governmental initiatives, they may benefit from greater flexibility and be easier to commence.
    A pilot with Spanish Universities and the European Quality Assurance Register (EQAR) are examples in this domain. 

    \subsubsection{Spanish Universities Pilot}
        Previously mentioned DI provider Gataca (Section \ref{sec:implementation_gataca}) is the lead partner in this pilot with numerous Spanish universities, namely Universidad Carlos III de Madrid (UC3M), Universidad de Murcia (UMU), and Universitat Rovira i Virgili (URV).  
        Gataca provides each university with a system to issue education credentials while utilizing the European Blockchain Service Infrastructure (EBSI) as the core blockchain \cite{214}. 
        The pilot aims to analyze the usability, impact, and potential of DI on wider implementations \cite{215}. 
        More recently (in 2024), Gataca has also partnered with the European Reform University Alliance (ERUA) to provide DI to more European universities. 
        Similarly to other pilots on DI, no reports are available. 

    \subsubsection{European Quality Assurance Register (EQAR)}
        The EQAR is a quality assurance agency for higher education in the EU. 
        Their SSI program enables the agency to issue diplomas digitally through the use of EBSI, DID, and VC standards.
        The initiative is based on the open-source SSI system, called walt.id \cite{236}, which was developed by a commercial company and supported by the EU. 
        The EQAR provides the SSI Kit, covering standard SSI capabilities and a wallet for users to store, manage, and share their identity data \cite{228}. 
        Users can selectively disclose their data to a verifier. 
        However, data export functionality is not mentioned and the wallet enables only import from third parties, which results in unclear credential portability \cite{228}. 

\subsection{Travel}
    Travel and border crossing are one of the popular examples of where DI is useful to adopt. 
    There were few attempts to deploy a system that allows people to pass through security and border control seamlessly, with the main goal of enabling traveling without a passport. 
    However, such an objective is difficult to achieve without compromising aviation security. 
    The known initiatives that utilize DI and SSI are the IATA Travel Pass and World Economic Forum (WEF) Known Traveler Digital ID. 

    \subsubsection{IATA Travel Pass}
        IATA Travel Pass was an application based on the Sovrin network and is currently no longer active. 
        It addressed the COVID-19 certificate presentation before a flight and test verification, intending to reduce paper-based documents fraud, while providing better data privacy and security to the certificate holder.
        \cite{46} points out that there was a security problem with the system, where certificate holders can be impersonated because their passport details are not verified. 
        Additionally, the certificates were issued through a web application managed by the company Evernym, which compromised decentralization benefits \cite{46}. 
    
    \subsubsection{WEF Known Traveler Digital ID (KTDI)}\label{sec:realworld_ktdi}
        The initiative aims to create a global digital ID for seamless travel that is based on a decentralized and interoperable identity platform. 
        The initiative pilot was recently resumed (in 2023) and allows travelers between the Netherlands and Canada to participate in the trial \cite{138}.
        The project involves a participating airline from each country, KLM and Air Canada, respectively, and technology provider companies Accenture, Vision Box and Idemia \cite{136}. 
        Accenture provides a blockchain and biometric technology to support the WEF initiative, classifying it as a distributed identity that aims to be interoperable with other identity systems \cite{137}. 
        The KTDI whitepaper refers to the system as a DI, as well as a traveler-centric concept.
        Based on the WEF specification paper, Hyperledger Indy is currently used for the pilot \cite{136}. 
        The primary purpose of this application still does not put users at the center of concern. 
        Multiple news articles \cite{139, 138} highlight the goal of the KTDI as "to speed up the process of screening travelers”. 

\section{Challenges}\label{sec:challenges}

The adoption of DI and SSI technologies faces many challenges from various perspectives. 
Largely, technological limitations continue to dominate the landscape, but with the majority of papers attempting to address a specific problem, such limitations require less attention. 
The social aspect continues to be largely unaddressed by the research community, with the usability and UX of decentralized applications still in their infancy. 
However, the largest overarching challenge that touches on all the smaller aspects, is the disconnect between proposed frameworks and real-world practices that results in a gap that obstructs practical deployment. 
This includes but is not limited to, the limited research that addresses the entire decentralized ecosystem, compatibility with current technologies, and absence of functional requirements for a DI and SSI. 
Therefore, there is a need to not only address the technological limitations of existing systems but also to understand the bigger picture of how the transition from traditional centralized identity management systems to a decentralized approach can occur. 

\subsection{Functional Requirements}\label{sec:challenge_requirements}
    Functional Requirements (FR) for DI or SSI frameworks are largely absent from the academic literature. 
    The existing systems base their architectures on principles developed by Christoper Allen \cite{61} and some authors use Cameroon's Laws of Identity \cite{198} for system comparison.
    As previously mentioned in Section \ref{sec:background_principles}, \cite{17} extends the initial set of SSI properties from 10 to 18. 
    The authors mention that some properties can be viewed as requirements that an SSI system should achieve and, thus, can be used to evaluate whether a system is an SSI or not \cite{17}.
    However, the mentioned properties continue to resemble Non-Functional Requirements (NFR), which do not provide a concrete basis for system analysis. 
    There have been very limited attempts to specify FR, with \cite{186} being the closest work. 
    \cite{186} provides a list of FRs for SSI, but the requirements are not mapped to the NFR that are extensively studied in \cite{61} and \cite{17}. 
    Additionally, the QuarkID whitepaper (Appendix II) \cite{181} maps SSI principles to design principles, which provide a remote resemblance to NFR. 
    However, the authors did not analyze or discuss those design principles in detail. 
    
    To achieve a solid foundation for SSI applications, it is necessary to map existing SSI principles to FR that an application can be based on. 
    Moreover, the FR are necessary for further evaluation of existing frameworks. 
    At the time of writing, evaluations of existing SSI systems are based on NFR, which results in limited reproducibility of the analysis and frequently unclear assignment of a principle to a framework.
    In other words, it is difficult, often impossible, to know why an author marks a solution as such (e.g., how did an author deduce that Sovrin's identity is interoperable or not?). 
    Therefore, the synthesis of FR is a crucial step in current landscape analysis that provides further instruction for real-world framework development and deployment. 

\subsection{Technological Limitations}\label{sec:challenges_tech}
    Despite being widely addressed in the academic literature, there are still technological challenges that limit the adoption and use of DI and SSI, including interoperability, portability and backward compatibility, key and VC recovery, scalability, offline verification, credential revocation, and metadata search in a blockchain. 

    \subsubsection{Interoperability, Portability \& Backwards Compatibility}\label{sec:challenge_tech_interop}
        The lack of interoperability between existing components and systems remains one of the important issues for the SSI adoption \cite{32, 45}. 
        Interoperability has five layers as defined by the National Interoperability Framework Observatory (NIFO) and European Telecommunications Standards Institute (ETSI): technical, syntactical, semantic, organizational, and legal \cite{27, 28, 33}. 
        However, technical interoperability proves to be the most challenging, as it involves the development of tools and architectures facilitating the exchange of data between different platforms \cite{33}. 
        \textit{Universal Resolver} (discussed in Section \ref{sec:standards_resolver}) has been designed to address the issue of interoperability \cite{19}. 
        However, the current solution is centralized and works as a trusted service between a client and blockchain, resulting in a need for a completely decentralized approach \cite{5, 22}. 
        Additionally, \cite{32, 44} point out that interoperability in real-world applications requires backward compatibility with existing systems. 
        To have a functional SSI system, the solution should consider bridging the new technology to the existing authentication and authorization solutions already in place \cite{49}. 
        \cite{102} describes a slightly different perspective, arguing that existing systems should be modified to facilitate SSI adoption. 
        
        Portability is another crucial aspect alongside interoperability. 
        The transfer of digital identity from one platform to another should be possible, timely, and smooth \cite{44}. 
        This is an important feature to ensure the continued existence of an identity in the case of a platform ceasing to exist. 
        However, there are some contradictory opinions about portability, with some authors arguing that an identity cannot be moved or copied from one platform to another and requires credential re-issuance \cite{49}, while others argue that DID design ensures portability without the need to reissue \cite{32} and wallets are the tool to ensure this \cite{17}. 
    
    \subsubsection{Key Recovery}
        Private key management represents another important issue in the adoption of SSI. 
        This challenge is composed of usability (discussed in Section \ref{sec:challange_ux}) and technological limitations. 
        Without a private key, an individual will not have an identity, and without a recovery mechanism in the case of key loss, the identity is not recoverable. 
        \cite{44} points out that the current implementations for key recovery and revocations rely on centralized servers or intermediaries. 
        Therefore, it is necessary to work toward the removal of such intermediaries to realize the full potential of the DI systems. 
        
        \cite{77} states that because SSI systems do not have a centralized authority, it is important to consider backup and recovery functionality in case any keys are lost. 
        The authors outlined two approaches to key recovery. 
        First, utilizing a seed phrase that creates deterministic keys based on mnemonic code. 
        Second, establishing a decentralized key management system (DKMS) that uses deterministic keys and seed phrases to enable social or offline recovery \cite{77}. 
        Yet, decentralized key recovery remains a challenge, with some systems not addressing this feature at all. 
        For example, Worldcoin (discussed in Section \ref{sec:implementations_worldcoin}), aims to provide every human with digital identity, but does not provide a recovery mechanism in case access is lost \cite{134}. 

    \subsubsection{Verifiable Claim Recovery}
        In addition to the key recovery challenge, a VC recovery mechanism may also be considered. 
        In one of the presentations given by \cite{63}, the author points out that there is not enough research done on this topic. 
        From the perspective of \cite{63}, credential recovery aims to guard against the possible disappearance of the issuer. 
        However, there is another interpretation of VC recovery presented by \cite{146}. 
        The authors propose a mechanism to recover device-bound anonymous credentials in case of device loss or change \cite{146}. 
        The problem addressed by \cite{146} borders closely with credential portability, which presents a significant problem. 
        Commonly, credentials issued are stored in wallet applications for easy management by the user. 
        However, only a few systems support credential and identity data export from the wallet, but no system addresses the portability of the data between wallets on different devices. 
        Therefore, the extensive reliance on wallet applications raises the question of what happens to identity data if the device is lost. 
        Considering today's frequency of upgrading mobile devices, this problem may be more widespread than considering only the case of a lost or stolen device. 
        Yet, the existing systems usually do not address this possibility. 
    
    \subsubsection{Scalability}
        The scalability and flexibility of a digital identity system are important for the adoption of new technologies \cite{44}.
        This is because some real-world applications may require a time-sensitive verification process (e.g., payment authorization or digital visa processing), as well as handling a large number of processes simultaneously. 
        Especially in the case of government digital identities, the system should operate with potentially millions of users, while maintaining its effectiveness. 
        Different systems target scalability differently. 
        For example, Sovrin utilizes two levels of nodes (one to accept write transactions and the other to observe nodes with read-only blockchain copies), while uPort was based on a public Ethereum network and did not address the question of scalability \cite{59}. 
        Thus, the scalability issue is not exclusive to the SSI and DI platforms, but inherited from the blockchain technology. 

    \subsubsection{Offline Verification}
        Since there are numerous situations when it is necessary to verify an identity or credential without internet access, offline verification is an important functionality to consider for feasible SSI adoption. 
        \cite{76} proposes an SSI architecture that allows offline authentication of a user. 
        The framework assumes that the authentication provider (or verifier) holds a copy of the blockchain locally. 
        Once the authentication is requested, the authentication provider needs to check whether the record is sufficiently recent \cite[p. 12]{76}. 
        However, the authors do not provide any guideline on what is “sufficiently recent” in this case. 
        Moreover, the reliance on a local copy of the entire blockchain may not be possible due to storage and processing limitations at the verifier. 
        \cite{79} points out that current SSI implementations largely depend on internet access for all operations. 
        Overall, there is limited research on how SSI functionalities, such as credential verification and revocation, can be performed offline. 

    \subsubsection{Credential Revocation}
        Credential revocation is an essential component of identity management \cite{paper}. 
        Not only the credentials may become invalid before their expiration date due to private key compromise or loss, but also some legal regulations require revocation to be implemented within a certain time period after the decision to revoke \cite{200}.
        A cryptographic accumulator is a widely used solution in DI revocation (e.g., Hyperledger Indy, Sovrin) that provides an efficient data structure whose size is independent of the number of revoked credentials. 
        However, a cryptographic accumulator has few limitations in terms of computational resources (e.g., for witness update), reliance on revocation authority in asymmetric variations, and the possibility of false positives in symmetric types (e.g., Bloom filters). 
    
    \subsubsection{Metadata Search}
        The problem of metadata data search is newly emerging for SSI systems but is more common to blockchain-based applications as a whole. 
        \cite{102} argues that there is a need for tools that will allow for efficient information search on the identity ledgers. 
        The metadata search problem relates to the search of a relevant schema previously published on the blockchain \cite{147}. 
        The authors defined schema in this context as a data structure for a specific domain \cite{147}. 
        For example, an HR company wants to receive resumes through an SSI network and needs to specify a schema for those resumes. 
        The company could either search the blockchain for previously published resume schema or create a new schema. 
        The authors argue that the second option is more expensive for the company than to reuse an existing schema \cite{147}. 

        Current approaches to this challenge frequently rely on storing blockchain transactions in an SQL database and then performing a regular database search \cite{147, 148}. 
        However, such an approach may not always be feasible because of the blockchain size and computing capabilities of the one who needs to perform the search. 
        There is a need for different approaches to facilitate blockchain-based search. 
    
\subsection{Social Considerations}
    Social considerations are the most critical to the real-world adoption of DI and SSI since these challenges deal with a human factor. 
    This section discusses often overlooked aspects of usability and user experience (UX), credential management, consent, and trust, highlighting the socioeconomic considerations and the need for more user-friendly and inclusive approaches. 

    \subsubsection{Usability \& User Experience}\label{sec:challange_ux}
        While technical features, security, and privacy concerns of digital identity providers are extensively explored and addressed in systems and academic literature, the mass adoption of DI and SSI still does not occur \cite{67, 45, 60}. 
        Several authors, such as \cite{67, 45, 60, 59, 102}, point out that UX, usability, and socioeconomic considerations are overlooked and rarely considered. 
        \cite{44} points out that usability research of identity management systems is in the developing stage and requires an improved long-term perspective. 
        
        Cryptographic key management and key recovery continue to be fundamental usability problems that are not only technically difficult but require different usability approaches and greater user education. 
        This is essential to ensure effective and safe wallet and key management by non-technical users \cite{44, 60}.
        A DI and SSI systems should be easy to understand and use by all categories of users irrespective of their knowledge and previous experience \cite{7}.
        
    \subsubsection{Self-Management}
        \cite{102} points out that because of poor usability, non-technical users may inadequately disclose information to verifiers. 
        From the other perspective, not all individuals can be entrusted to self-manage their identities and decide when and with whom to share which data \cite{39}.  
        Additionally, a sophisticated and demanding key management system is disadvantageous to various social groups such as the elderly and disabled \cite{166}. 
        Therefore, there should be a mechanism to determine whether a user should have complete control over their identity data, based on some external factors (e.g., medical conditions). 
        At the time of writing, no author discussed whether such a mechanism would undermine the concepts of an SSI and whether such a beneficiary role is technically and conceptually feasible. 
        \cite{159} points out that Sovrin has the concept of a "guardian", that addresses this issue by “handling an identity on behalf of a vulnerable person or anyone else incapable of managing their digital wallet“. 
        However, the authors did not provide a detailed discussion on how this is achieved in the mentioned system. 
        The QuarkID framework recognizes this challenge in their whitepaper \cite{181}. 
        However, the only relevant document QuarkID refers to is the Sovrin Foundation's concept of guardianship. 

    \subsubsection{Consent}\label{sec:challenge_social_consent}
        The issue of self-management can be extended further to the problem of consent. 
        DI initiatives offered on a governmental level raise the possibility of forcing the system on participants, while not addressing those who do not wish to join the system. 
        This raises the possibility of total exclusion of some people, not only from the services but from certain life processes. 
        Moreover, this problem does not only concern those individuals who completely refuse to participate in a new digital identity initiative but also those who are not willing to share all of the required information for a service. 
        A service provider or verifier still has a superior position over the data holder, as they can dictate what information they require to be shared with them to use their service. 
        A service provider may request more data than they need, and the user would be forced to either comply or be excluded. 
        In the situation where similar services are provided by more than one provider, this is manageable, but in the case of governmental institutions or international travel, there is no choice. 
        \cite{128} points out that the disclosure of personal information is a prerequisite in several cases (e.g., international travel) with or without a specific application. 
        According to the author, the use of digital identity does not endanger the user’s privacy but allows greater cooperation between different parties \cite{128}. 
        
        Not only is exclusion possible due to an unwillingness to share required information, but people who do not have a suitable device to run a wallet application would also be in the risk category \cite{7}. 
        Thus, such individuals should be able to store credentials on physical items, such as identity cards, to be able to use a service \cite{7}. 

        Moreover, \cite{84} points out another perspective on the consent challenge an SSI faces. 
        The authors argue that user consent to privacy and data sharing notice is not easy to implement but mandated by the regulations (e.g., GDPR), which leads to "consent fatigue" where a user is continuously required to respond to privacy notifications \cite{84}.

    \subsubsection{Trust}\label{sec:challenge_trust}
        The notion of SSI is shaped by the interests of social actors \cite{39}. 
        \cite{158} raises concerns over how the term SSI is currently used by companies and government-backed initiatives. 
        According to the authors, there is an ambiguity and lack of common understanding of an SSI, leading to a large number of companies using the term SSI for marketing purposes, even when their systems do not involve these features \cite{158}. 
        Thus, it might be difficult to convey a message about the benefits of DI to the end users.
        Many still do not recognize the privacy benefits such a system can provide for a specific domain, resulting in a \textit{communication challenge} and \textit{lack of trust} \cite{60, 158}. 
        Additionally, trust in verifiers and other parties is difficult to achieve because there is no feasible way to quantify it \cite{44}. 
        The transparent flow of data, confidentiality, integrity, authenticity, non-repudiation, and robustness of a system should be demonstrated to its users to ensure sufficient trust \cite{44}. 
        \cite{170} points out the importance of UX to facilitate trust, particularly when the user does not have alternative options (e.g., government services). 

        Based on interview results conducted by \cite{32}, the authors outline that the feasibility of SSI adoption largely depends on the purpose of the usage. 
        Furthermore, the domain influences the social perception and understanding of SSI \cite{39}.
        For example, \cite{39} points out the difference between Europe and the Anglosphere (e.g., United States, Canada) in the perception of trust in authority and the view on SSI. 
        With Europe's perception of centralized management as the “source of truth”, the SSI potential is emphasized in the “government’s duties as a supervisor over citizens” \cite[p.2550]{39}. 
        In contrast, the Anglosphere tends to view SSI as a potential to re-establish trust in centralized institutions \cite{39}. 
        Moreover, from a political perception, the name "self-sovereign identity" may be associated with controversial politics and emphasize some refusal to acknowledge government authority \cite{158}. 
        \cite{158} points out that it is still not clear how such an association could affect the wide adoption of SSI technology. 

\subsection{Economic Considerations}\label{sec:challenge_economic}        
    The economic considerations are typically overlooked in the design of DI and SSI systems. 
    Nevertheless, this is a determining factor on whether a novel system will become widely adopted in real-world interactions. 
    \cite{87} points out that the user is unlikely to migrate their identity data to an SSI system unless the new technology provides a superior functionality or is economically better than the existing alternative. 
    Notwithstanding, DI and SSI have the potential to offer greater benefits to a user, but such benefits and overall utility and costs have to be properly communicated to the user. 
    While the benefits to a user are clearly established in the research community, the costs still require further investigation, similar to the analysis of decentralized storage and comparison to a centralized system provided by \cite{240}.

\subsection{Reliance on Private Sector}
    A relatively large number of real-world implementations rely on systems that are designed, managed, and owned by private companies, corporations, or start-ups, which results in a lack of governance decentralization (discussed in the previous section). 
    Such companies are primarily business-oriented and provide a competitive user-friendly system, but in return introduce a large dependency on that software provider \cite{7}. 
    Private sector companies are usually closed source, which results in limited accessibility, flexibility, and the absence of community collaborations.
    Meanwhile, systems offered as a Software-as-a-Service (SaaS) behind a paywall offer guaranteed support, easier deployment, and software complexity management \cite{231}. 
    The benefits of adopting SaaS solutions come as significantly higher financial costs and potential vendor lock-in \cite{231}.
    Additionally, as was observed with a few examples, such as uPort and Jolocom, company-backed systems do become inactive for a variety of reasons. 
    However, it is not clear how this impacts ongoing pilots or integration due to the lack of reporting by service providers and users. 
    To enable successful real-world integration of DI or SSI, there must be no reliance on individual companies. 
    Such an approach would facilitate better reliability and provide users with functionality that benefits the user rather than the business. 

\subsection{Security}\label{sec:challenge_security}
    Despite the DI or SSI approaches being more secure than reliance on centralized databases, security is still a crucial consideration for the design of a new digital identity approach. 
    The security problem is addressed differently in the case of DI and SSI, placing a user at the center of consideration, instead of a malicious actor. 
    Thus, shifting the security concern on the interactions between a user and a system. 
    For instance, the main cause of information loss in 2024 was due to non-cyber incidents, comprising 71\% of the total incidents reported \cite{242}. 
    Information Commissioner's Office of the UK defines non-cyber incidents as "a type of breach that does not have a clear online or technological element which involves a third party with malicious intent" \cite{243}.
    That is, the user was responsible for an incident, such as accidentally emailing information to the wrong recipient. 
    This raises a concern about whether users can securely manage their private information and avoid its loss in practice. 
    This problem leads back to the usability and UX challenges discussed in Section \ref{sec:challange_ux}. 
    Therefore, these two challenges should have a security consideration to produce not only usable systems but also secure ones. 

\subsection{Privacy}
    The question of user privacy in DI and SSI systems is not entirely clear. 
    From one point of view, minimal disclosure of credential data enhances privacy, reducing the risk of unnecessary data sharing \cite{231}.
    \cite{231} points out that the reliance on the presentation of a credential and its storage at the owner, instead of the verifier, results in improved compliance with data protection regulations. 
    However, credentials linkability should not be dismissed.  
    For instance, \cite{76} points out that the reliance on the credential revocation approach where credentials are posted to the blockchain leads to user identification. 
    Particularly, during the authentication process, the credential issuer may link verifiers and observe requests sent by a user. 
    
    Furthermore, metadata privacy has not established itself as a critical challenge yet, as there is no clear and official guidance on whether metadata is considered private data and requires the same level of protection. 
    Metadata, defined as data about data, is most critical within VCs and includes information such as the DID of the issuer, issuance date, validity period, and type \cite{91, 229}. 
    With the increasing adoption of DI and SSI and the development of privacy regulations (e.g., General Data Protection Regulation (GDPR)), there may be a need to address the privacy of metadata to ensure systems remain privacy-preserving for a long period of time. 
    Moreover, none of the DI systems based on public blockchains address the deletion of metadata from the ledger \cite{12}. 
    If metadata is eventually classified as private data and falls under GDPR, its removal from blockchains will become a critical issue. 
    
\subsection{Standards}
    DI is still facing standardization challenges \cite{60}. 
    As mentioned in Section \ref{sec:standards}, the main standards body W3C Verifiable Credentials Working Group released a stable version of DID and VC standards only in 2022, and the refined version in 2024. 
    Immature standards in turn lead to limited interoperability and portability of the credentials and identities, risking locking users to a particular identity provider without the opportunity to move data between systems or use it interchangeably between systems \cite{60}. 
    \cite{39} points out that the view on SSI is more definite and defined when the initiative is supported by the European Commission.
    
    As shown in Tables \ref{tab:implementations_ssi}, \ref{tab:implementations_non_ssi} and \ref{tab:real_world}, the majority of the systems and real-world implementations adopt W3C standards. 
    Yet, some components of the W3C standardization require further development, such as a Universal Resolver (mentioned in Section \ref{sec:standards_resolver}) that, at the time of writing, relies on a centralized architecture. 

\subsection{Regulations}\label{sec:challenges_regulations}
    Regulations present a challenge in any blockchain-based technology, as there is a need for regulations that specify the rules clearly for blockchain use cases \cite{12}.
    Moreover, there is uncertainty in the legal and regulatory frameworks that concern DI and SSI \cite{32}.
    Even though an SSI system is usually based on a distributed platform and distributed globally, in most cases, its operations are restricted by the local regulations where the network operates \cite{77}. 
    eIDAS is one of the common regulatory initiatives in the European Union that addresses digital identity. 
    One of the strengths of this regulation is the promotion of the SSI model. 
    However, \cite{172} and \cite{171} point out that under the newly updated Article 45 in the eIDAS legislation, the EU could monitor the online interactions of SSI wallet owners.
    Therefore, as \cite{77} points out, regulations would influence the technical design choice of the entire SSI platform to comply with laws.
    
    Additionally, privacy laws, such as GDPR, require that users be informed as to why their data is collected, how it will be used, who will be processing it, where it will be transferred, how users can erase it, and how they can stop processing \cite{134}. 
    Not only may this be challenging technically, but this also represents a usability problem. 
    Terms and conditions and privacy notices are often designed for compliance rather than user understanding, resulting in lengthy, confusing, and non-informative documents that most users do not read, leading to "consent fatigue" (as mentioned in Section \ref{sec:challenge_social_consent}). 
    For example, Worldcoin provides terms and conditions and privacy notices on at least five separate web pages \cite{140}. 
    
\subsection{Governance}\label{sec:challenge_governance}
    \cite{76} defines governance as the management of the system organization and participation in the consensus mechanism of the underlying blockchain. 
    \cite{32} points out that in SSI, the interactions between participants (e.g., issuers and verifiers) are managed by a governance authority. 
    According to the authors, a governance authority is compromised by any number of issuers and they oversee the governance framework and rules, such as business, legal, and technical guidelines \cite{32}.
    However, the authors do not address the case of completely decentralized SSI, where there is no governance authority present to collect and publish regulatory guidelines. 

    Importantly, as discussed in Section \ref{sec:background_technical}, for a system to be considered an SSI, the decentralization of governance is a determining factor. 
    In other words, no single organization or authority has the power over system components. 
    However, with a closer examination of Table \ref{tab:real_world}, one can observe that 6 (or 7, considering QuarkID) out of 10 (60\%) real-world examples rely on private instances of the blockchain, which assumes the governance of the system is under control of an organization.
    At the same time, systems outlined in Tables \ref{tab:implementations_ssi} and \ref{tab:implementations_non_ssi} largely lack widespread real-world deployments, and at least 11 out of 20 (55\%) utilize public blockchain networks, enabling a complete decentralization of governance. 
    Thus, achieving a decentralization of governance in real-world SSI deployments is challenging to achieve.

\section{Conclusion}\label{sec:conclusion}

By expanding the current state of the art of the DI and SSI technology, this paper identifies challenges that have to be addressed to facilitate a transition toward a user-focused identity management model. 
Furthermore, this paper outlines why a DI and SSI are not widely used in everyday interactions yet.
This section summarizes reasons for slow adoption and provides recommendations for future research. 

\subsection{Make the User Aware of the Technology}
    As mentioned in Section \ref{sec:challenge_economic}, the user is unlikely to adopt a new technology unless it provides a superior functionality or is economically better than the existing alternative. 
    However, as discussed in Section \ref{sec:challenge_trust}, since the notion of SSI is ambiguous and frequently used for marketing purposes, it may be difficult to convey an accurate and convincing message to potential users about the benefits of the decentralized approach to identity management. 
    Therefore, the mitigation of this challenge is manifold, with the first step being to 
    promote clear and distinct definitions of DI and SSI (summary of definitions is provided in Section \ref{sec:background_definitions} and elaborated in greater detail in \cite{chapter}).
    Building on a better understanding of both concepts, it becomes apparent that a pure SSI system is much harder to adopt in practice than a DI since the decentralization of governance is rarely achieved in real-world deployments (discussed in Section \ref{sec:challenge_governance}).
    Moreover, researchers and developers need a practical framework to evaluate existing implementations and arrive at a concrete conclusion about whether it is a DI or an SSI. 
    A practical framework needs to build on the existing SSI NFR, develop FR, and provide a meaningful mapping between the two (this is elaborated further in Section \ref{sec:challenge_requirements}. 
    Such a framework would extend the knowledge of an application beyond the current reliance on academic sources, official websites, and documentation.
    That is, certain actions could be performed with an application and based on the output, solid and reproducible conclusions could be reached about the NFR of a system.
    Thus, providing a better understanding of what works in the real-world and what can be categorized as an SSI. 
    With a clear understanding of the notions and how existing systems are categorized, an accurate message about the differences and benefits of a system can be conveyed and communicated to the users. 
    Improving user awareness of the underlying technology is the crucial step to facilitate the acceptance and utility of DI and SSI.  
    Additionally, a greater connection between research and real-world use cases should be achieved, including the need for consistent reporting on the conducted pilots and already deployed systems. 
    Therefore, improving communication and trust, and ensuring the user is aware of the underlying technology that they are using.

\subsection{Make the Decentralized Application Usable}
    The ability to foster user trust in new technology and communicate provided functionalities may require a multi-dimensional approach, with a significant dimension being usability and UX. 
    Since the user only interacts with an application that represents all the functionalities of a system, a transparent, clear, and, at the same time, informative user interface is non-negotiable. 
    This pressure for flawless usability advances further in DI and SSI applications, where the user is the central focus of a system.
    The decentralization aspect and its benefits should be communicated unambiguously to the user through the application interface, enabling a clear differentiation between decentralized and centralized identity management approaches, and awareness of which the system utilizes.
    Moreover, as mentioned in Section \ref{sec:challange_ux}, a DI system itself includes unique functionalities, such as cryptographic key management and recovery, that are not present at all in the centralized approaches.
    Security is no longer assessed with only an attacker in mind, but with the user at the center of concern, since he has all the capabilities to store and share his personal data (discussed in Section \ref{sec:challenge_security}. 
    A complicated application interface and lack of user understanding about the application's functionalities may lead to flawed credentials sharing, exposing sensitive data that should remain inaccessible. 
    Additionally, the need for regulatory compliance and privacy notices extends the communication challenge.
    These should be easy to understand and communicated effectively to the users to reduce consent fatigue (as outlined in Section \ref{sec:challenges_regulations}) and promote improved perception of the privacy benefits a DI and SSI offer. 
    Therefore, there is a need for more research on the usability and UX of DI and SSI applications, as well as addressing the awareness of the benefits of a DI and SSI by a user.

\subsection{Make the Personal Data Universal}
    Even though isolated technological limitations do not represent the complete landscape of challenges to DI and SSI adoption, these are still important to address to make real-world adoption feasible and widespread.
    Interoperability, portability, and backward compatibility comprise the most limiting factor of adoption since it is not practical to adopt a system that can only be used in isolation from existing conventional technologies and other systems.
    As discussed in Section \ref{sec:challenge_tech_interop}, there is a need for backward compatibility to ensure new identity management can work together with conventional systems already in place.
    Moreover, credentials and personal data should be interoperable and portable between the user's devices, without the need to reissue. 
    To achieve this, the current centralized approach (the Universal Resolver, as mentioned in Section \ref{sec:challenge_tech_interop}) should see greater decentralization attempts, and standards (outlined in Section \ref{sec:standards}) finalized and promoted to wider adoption. 
    Similarly to the ambiguity of the DI and SSI notions, interoperability and portability are not always defined in the same fashion. 
    For instance, there is an inconsistency in understanding portability and how it is achieved in a decentralized setting, with some authors (e.g., \cite{17}) considering the application to be solely responsible for it.
    There is a need for a better understanding of the application's role in ensuring portability and how the disappearance of a system and its native wallet impacts the credentials and user data.

\section*{Acknowledgment}

\noindent We express our gratitude to Timothy Arney for proofreading this work and making it better.


\bibliographystyle{IEEEtran}
\balance
\small
\bibliography{main}

\end{document}